\documentclass[journal, 12pt, draftclsnofoot, onecolumn]{IEEEtran}
\IEEEoverridecommandlockouts
\def\BibTeX{{\rm B\kern-.05em{\sc i\kern-.025em b}\kern-.08em
    T\kern-.1667em\lower.7ex\hbox{E}\kern-.125emX}}

\usepackage{amsmath,amssymb,amsfonts,graphicx,mathtools,nccmath,amsthm,float}
\usepackage{algorithm,multicol,multirow,tikz,cite,array,booktabs,url}
\usepackage{algpseudocode}
\usepackage{cite}
\usepackage{textcomp}
\usepackage{xcolor}
\usepackage[inline]{enumitem}
\usepackage{subcaption}
\usepackage[bookmarks,colorlinks]{hyperref}

\usepackage[export]{adjustbox}

\graphicspath{{figures/}}
\DeclareGraphicsExtensions{.pdf}

\newcommand{\vect}[1]{\boldsymbol{\mathbf{#1}}}

\newtheorem{Thm}{Theorem}
\newtheorem{Cor}{Corollary}
\newtheorem{Lem}{Lemma}
\newtheorem{Prop}{Proposition}

\DeclareMathOperator{\diag}{diag}

\DeclareMathOperator{\trace}{Tr}
\DeclarePairedDelimiter{\norm}{\lVert}{\rVert}

\algnewcommand\algorithmicinput{\textbf{Set}}
\algnewcommand\Set{\item[\algorithmicinput]}
\algnewcommand\algorithmicinitial{\textbf{Initialize}}
\algnewcommand\Initialize{\item[\algorithmicinitial]}

\let\oldReturn\Return
\renewcommand{\Return}{\State\oldReturn}

\begin{document}

\title{Counteracting Eavesdropper Attacks Through Reconfigurable Intelligent Surfaces: A New Threat Model and Secrecy Rate Optimization}
\author{George~C.~Alexandropoulos,~\IEEEmembership{Senior Member,~IEEE,} Konstantinos~D.~Katsanos,~\IEEEmembership{Graduate Student Member,~IEEE,} Miaowen Wen,~\IEEEmembership{Senior Member,~IEEE,} and Daniel B. da Costa,~\IEEEmembership{Senior Member,~IEEE}
\thanks{Part of this paper was presented in \textit{IEEE ICC}, Montreal Canada, 14-23 June 2021 \cite{Conf}. This work has been supported in part by the EU H2020 RISE-6G project under grant number 101017011 and in part by the Hellenic Foundation for Research and Innovation (HFRI) and in part by the General Secretariat for Research and Technology (GSRT), through the HFRI Ph.D. Fellowship Grant GA under Grant 1392.}
\thanks{G.~C.~Alexandropoulos and K.~D.~Katsanos are with the Department of Informatics and Telecommunications, National and Kapodistrian University of Athens, Panepistimiopolis Ilissia, 15784 Athens, Greece (e-mails: \{alexandg, kkatsan\}@di.uoa.gr).}
\thanks{M. Wen is with the School of Electronic and Information Engineering, South China University of Technology, 510641 Guangzhou, China (email: eemwwen@scut.edu.cn).}
\thanks{D. B. da Costa is with the AI and Digital Science Research Center, Technology Innovation Institute, 9639 Masdar City, Abu Dhabi, United Arab Emirates (e-mail: danielbcosta@ieee.org).}
}

\maketitle
\begin{abstract} 
The potential of Reconfigurable Intelligent Surfaces (RISs) for energy-efficient and performance-boosted wireless communications is recently gaining remarkable research attention, motivating their consideration for various $5$-th Generation (5G) Advanced and beyond applications. In this paper, we consider a Multiple-Input Multiple-Output (MIMO) Physical Layer Security (PLS) system with multiple data streams including one legitimate passive RIS and one malicious passive RIS, with the former being transparent to the multi-antenna eavesdropper and the latter's presence being unknown at the legitimate multi-antenna transceivers. We first present a novel threat model for the RIS-boosted eavesdropping system and design a joint optimization framework for the eavesdropper's receive combining matrix and the reflection coefficients of the malicious RIS. Focusing next on the secrecy rate maximization problem, we present an RIS-empowered PLS scheme that jointly designs the legitimate precoding matrix and number of data streams, the Artificial Noise (AN) covariance matrix, the receive combining matrix, and the reflection coefficients of the legitimate RIS. The proposed optimization algorithms, whose convergence to at least local optimum points is proved, are based on alternating maximization, minorization-maximization, and manifold optimization, including semi-closed form expressions for the optimization variables. Our extensive simulation results for two representative system setups reveal that, in the absence of a legitimate RIS, transceiver spatial filtering and AN are incapable of offering non-zero secrecy rates, even for malicious RISs with small numbers of elements. However, when an $L$-element legitimate RIS is deployed, confidential communication can be safeguarded against eavesdropping systems possessing even more than a $5L$-element malicious RIS. 
\end{abstract}
\begin{IEEEkeywords}
Artificial noise, area of influence, MIMO, manifold optimization, physical layer security, reconfigurable intelligent surface, threat modeling, secrecy rate.
\end{IEEEkeywords}

\section{Introduction} \label{Sec:Intro} 
Reconfigurable Intelligent Surfaces (RISs) have been recently envisioned as a revolutionary means to transform any passive wireless communication environment to an active reconfigurable one \cite{George_RIS_TWC2019, Wu_RIS_TWC2019,rise6g}, offering increased environmental intelligence for diverse communication objectives. An RIS is an artificial planar structure with integrated electronic circuits \cite{Kaina_metasurfaces_2014} that can be programmed to manipulate an incoming electromagnetic field in a wide variety of functionalities \cite{Marco_Visionary_2019, HMIMO,Tsinghua_RIS_Tutorial}. Among the various RIS-enabled objectives belongs the Physical Layer Security (PLS) \cite{Yang_ComMag_2015}, which is considered as a companion technology to conventional cryptography, targeting at significantly enhancing the quality of secure communication in beyond $5$-th Generation (5G) wireless networks. Recent theoretical comparisons between RISs and conventional relaying schemes (such as decode-and-forward and amplify-and-forward), in terms of the achievable average secrecy capacity, have witnessed the increased potential of RIS-empowered legitimate systems under eavesdropping attacks \cite{Mensi_2021}.

One of the very first recent studies on RIS-enabled PLS systems is \cite{Chen_2019}, which considered a legitimate Multiple-Input Single-Output (MISO) broadcast system, multiple eavesdroppers, and one RIS for various configurations of the reflection coefficients of its discrete unit elements. In that work, aiming at safeguarding legitimate communication, an Alternating Optimization (AO) approach for designing the RIS phase configuration matrix and the legitimate precoder was presented together with a suboptimal scheme based on Zero Forcing (ZF) precoding that nulls information leakage to the eavesdroppers.
Recently in \cite{Hong_2020}, an RIS-assisted Multiple-Input Multiple-Output (MIMO) PLS system was considered, where the precoding matrix for fixed number of data streams, the Artificial Noise (AN), and the RIS reflection configuration of the legitimate side were jointly designed targeting the secrecy rate maximization. The RIS-aided Gaussian MIMO wiretap channel was also investigated in \cite{Dong_2020b}, assuming both full and no knowledge of the eavesdropper's Channel State Information (CSI). Therein, the transmit covariance and the RIS phase configuration matrices were jointly optimized in order to further enhance the achievable secrecy rate. In \cite{Feng_2019}, the energy consumption problem under secrecy rate guarantees was investigated for the MISO RIS-aided wiretap channel, assuming both perfect and statistical eavesdropper's CSI knowledge, revealing the RIS deployment benefits. Aiming at minimizing the same performance metric, when the eavesdropper's CSI is unavailable, an RIS-based design was proposed in \cite{Wang_2020b}, which was combined with a jamming strategy to further enhance secrecy. Based on a statistical CSI error model for the RIS-parameterized cascaded channel, 
the power minimization problem for a system with multiple single-antenna eavesdroppers was considered in \cite{Hong_2021b}, where AN was utilized to boost the secrecy performance. Exploiting statistical CSI for a single-antenna eavesdropper, the ergodic secrecy rate maximization problem was studied in \cite{Liu_2021f}.  
The energy efficiency maximization of a MISO system with multiple legitimate Receivers (RX) and eavesdroppers, was examined in \cite{Li_2022c} assuming CSI uncertainty at the transmitter with respect to the unintended users. By imposing outage probabilistic constraints via the considered imperfect CSI model, it was shown that the proposed scheme achieves improved energy efficiency when the RIS ignores CSI uncertainty. The transmission of Confidential Bit Streams (CBSs) to the legitimate RX in the presence of a multi-antenna eavesdropper with the help of an RIS was studied in \cite{Shu_2021b}, where it was numerically shown that the secrecy rates can be even doubled when using two CBSs compared to a single one. 

On the other hand, any promising physical-layer technology, like MIMO and RISs, can be maliciously deployed from the eavesdropping side. A new type of attack, termed as RIS jamming attack, was investigated in \cite{Lyu_2020}, according to which a passive RIS reflects jamming signals harming legitimate communications. The presented simulation results exhibited that the legitimate received signal can be downgraded up to $99\%$, witnessing that an RIS can be effectively used by the eavesdropping side for zero-power jamming. A malicious RIS was also considered in \cite{Zheng_2020} on the uplink, which was designed to reflect pilot or data signals from legitimate users intended for the legitimate Base Station (BS). 
Leveraging the benefits of an RIS, a novel pilot contamination attack for the eavesdropper who controls an RIS was proposed in \cite{Huang_2021}. As countermeasure, a sequential detection scheme from the BS side, combined with a cooperative channel estimation protocol,
was presented to reduce information leakage and enable secure transmissions. 

All above recent studies corroborate that RISs are capable to offer increased safeguarding flexibility for legitimate systems, but they can be also maliciously adopted by non-legitimate systems for eavesdropping in a transparent and energy efficient manner. 
In this paper, we consider a multi-stream MIMO legitimate system operating in the vicinity of a multi-antenna eavesdropping system, where each side deploys a passive RIS which is transparent to the other system. 
Differently from our recent work \cite{Conf}, where perfect CSI was considered at both the legitimate and eavesdropping parts, we assume statistical CSI knowledge with respect to the eavesdropping links at the legitimate system, while partial CSI knowledge is available at the eavesdropping system. Focusing first on the latter, we present a joint design framework for the eavesdropper's combining matrix and the reflection coefficients of the malicious RIS. Then, by formulating and solving a novel joint design problem for the legitimate system, whose objective is the secrecy rate maximization, we present an RIS-empowered PLS scheme incorporating legitimate precoding and AN, receive combining, and reflective beamforming from the legitimate RIS. The proposed framework for the legitimate system explicitly optimizes the number of data streams. Our extensive simulation results demonstrate that our RIS-empowered PLS scheme can secure confidential communication even in the presence of a malicious RIS with large numbers of elements. The main contributions of this paper are summarized as follows.

\begin{itemize}
 \item We consider a new threat model according to which the legitimate system is unaware of the existence of a malicious RIS and the RIS-boosted eavesdropping system ignores the presence of a legitimate RIS. Based on this model, we design the receive combining matrix of the multi-antenna eavesdropper and the reflection coefficients of the malicious RIS.
 \item Focusing on the RIS-empowered MIMO legitimate system, we present a novel joint design of the legitimate precoding matrix and number of data streams, the AN covariance matrix, the receive combining matrix, and the reflective beamforming from the legitimate RIS.
 \item The proposed design algorithms are mainly based on alternating maximization, minorization-maximization (MM), and Manifold Optimization (MO), including semi-closed form expressions for the main optimization design variables. 
 We additionally prove the convergence of the presented algorithms to at least locally optimal points.
 \item 
 Considering two representative setups for the legitimate RIS placement, we investigate the average number of data streams per transmit power level and introduce the Area of Influence (AoI) of the proposed PLS scheme, which characterizes the achievable secrecy rate in a given geographical area. 
\end{itemize}

\textit{Notations:} Vectors and matrices are denoted by boldface lowercase and boldface capital letters, respectively. The transpose, conjugate, Hermitian transpose, inverse and pseudo-inverse of $\mathbf{A}$ are denoted by $\mathbf{A}^T$, $\mathbf{A}^*$, $\mathbf{A}^H$, $\mathbf{A}^{-1}$, and $\mathbf{A}^{\dagger}$ respectively, and $|\mathbf{A}|$ is the determinant of $\mathbf{A}$, while $\mathbf{I}_{n}$ ($n\geq2$) and $\mathbf{0}_{m \times n}$ are the $n\times n$ identity matrix and the $m\times n$ zeros' matrix, respectively. ${\rm Tr}(\mathbf{A})$ and $\norm{\mathbf{A}}_F$ represent $\mathbf{A}$'s trace and Frobenius norm, respectively, while notation $\mathbf{A}\succ\vect{0}$ ($\mathbf{A}\succeq\vect{0}$) means that the square matrix $\mathbf{A}$ is Hermitian positive definite (semi-definite). $[\mathbf{A}]_{i,j}$ is the $(i,j)$-th element of $\mathbf{A}$, $[\mathbf{a}]_i$ is $\mathbf{a}$'s $i$-th element, and ${\rm diag}\{\mathbf{a}\}$ denotes a square diagonal matrix with $\mathbf{a}$'s elements in its main diagonal. $\odot$ and $\otimes$ stand for the Hadamard and Kronecker products, respectively, while $\operatorname{vec}(\vect{A})$ indicates the vector which is comprised by stacking the columns of $\vect{A}$, and $\operatorname{vec}_{\rm d}(\vect{A})$ denotes the vector obtained by the diagonal elements of the square matrix $\vect{A}$. $\operatorname{unit}(\mathbf{a})$ means $\mathbf{a}$ has its elements normalized, while $\nabla_{\mathbf{a}}f$ and $\nabla_{\mathbf{a}}^{\rm R}f$ denote, respectively, the Euclidean and the Riemannian gradient vectors of a scalar function $f$ along the direction indicated by $\mathbf{a}$. $\mathbb{C}$ represents the complex number set, $\jmath \triangleq \sqrt{-1}$ is the imaginary unit, $|a|$ denotes the amplitude of the complex scalar $a$ and $\Re(a)$ its real part. $\mathbb{E}[\cdot]$ is the expectation operator. $\mathbf{x}\sim\mathcal{CN}(\mathbf{a},\mathbf{A})$ indicates a complex Gaussian random vector with mean $\mathbf{a}$ and covariance matrix $\mathbf{A}$. $\mathcal{O}(f(x))$ represents the Big-O notation for the function $f(x)$.

\section{System and Signal Models} \label{Sec:Sys_Model}
\begin{figure}[t!]
\centering
\includegraphics[scale=0.5]{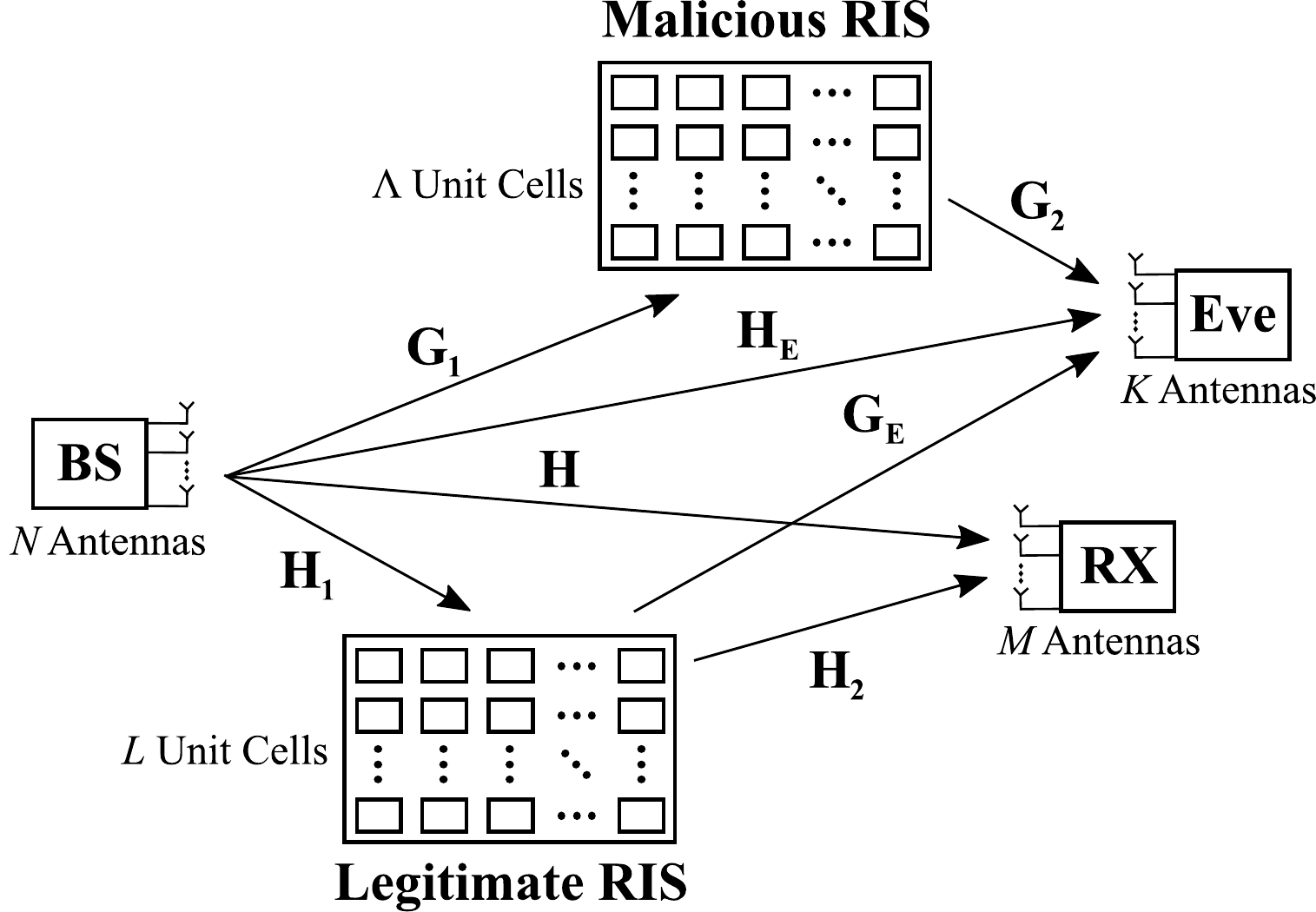}
\caption{\small{The considered RIS-empowered PLS communication system comprising three multi-antenna nodes and two RISs, one serving the eavesdropper Eve and the other the legitimate BS-RX communication. The BS is assumed unaware of the existence of the malicious RIS, the same is assumed for Eve regarding the legitimate RIS's existence.}}
\label{fig:System_Model}
\end{figure}

The system model in Fig$.$~\ref{fig:System_Model} consists of an $N$-antenna legitimate BS aiming to communicate with a legitimate RX having $M$ antennas. This downlink MIMO transmission is assumed to be further empowered by a legitimate RIS with $L$ reflecting elements, which is placed either close to the BS or RX. In the vicinity of the legitimate BS-RX link exists a $K$-antenna ($K\geq M$) eavesdropper (Eve) whose overhearing is assisted by an RIS having $\Lambda$ reflecting elements.

\subsection{Proposed Threat Model and PLS Operation} \label{Sec:Threat_Model}
The $\Lambda$-element malicious RIS is assumed to be located close to Eve and in the line-of-sight (LOS) from the legitimate BS to facilitate legitimate information decoding.
We assume that the legitimate RIS is connected to the legitimate node via dedicated hardware and control signaling for online reconfigurability \cite{Tsinghua_RIS_Tutorial}; the same holds between Eve and the malicious RIS. The BS knows about the existence of Eve and focuses on securing its confidential link with RX; however, it is unaware of the presence of the malicious RIS. The deployment of the legitimate RIS by the legitimate BS-RX communication is assumed transparent to Eve.

By deploying pilot-assisted CSI estimation \cite{Tsinghua_RIS_Tutorial}, we assume that the BS possesses perfectly the channel matrices $\vect{H}\in\mathbb{C}^{M \times N}$, $\vect{H}_1\in\mathbb{C}^{L \times N}$, and $\vect{H}_2\in\mathbb{C}^{M \times L}$, referring to the wireless links from the BS to RX, BS to legitimate RIS, and legitimate RIS to RX, respectively.
For the channel matrix $\vect{H}_{\rm E}\in\mathbb{C}^{K \times N}$ between the BS and Eve and the matrix $\vect{G}_{\rm E} \in \mathbb{C}^{K \times L}$ for the channel between the legitimate RIS and Eve, we adopt the statistical CSI model of \cite{Wang_2021}, according to which the up to the second-order statistics of the channels are available at the BS.
Recall that the BS is unaware of the existence of the malicious RIS, hence, it has no knowledge of the channel $\vect{G}_1\in\mathbb{C}^{\Lambda \times N}$ between itself and that RIS, neither of the malicious RIS to Eve channel $\vect{G}_2\in\mathbb{C}^{K \times \Lambda}$. On the other hand, Eve is assumed to perfectly possess $\vect{G}_2$ (e.g., via using any of the receiving RIS hardware architectures \cite{George_RIS_2020,HRIS_Mag} and pilot-assisted channel estimation), while, capitalizing on the assumption that the malicious RIS is located in the LOS from the BS, it can partially estimate the channel matrix $\vect{G}_1$.
We finally assume that there is no actual channel between the malicious RIS and RX; the extension of the optimization framework in this paper to this more general case is straightforward and is left for future investigation.

\subsection{Received Signals and Secrecy Rate} \label{Sec:Signal_Model} 
To secure the confidentiality of the legitimate BS-RX link, the BS applies AN \cite{Liu_2017} that is jointly designed with the BS precoding scheme and the legitimate RIS phase configuration vector $\vect{\phi}\triangleq[e^{\jmath \theta_1}\,\,e^{\jmath \theta_2}\,\,\cdots\,\,e^{\jmath \theta_L} ]^T\in\mathbb{C}^{L \times 1}$, where $\theta_{\ell}$ with $\ell = 1,2,\ldots,L$ denotes the phase shifting value at the $\ell$-th RIS unit element. We represent by $\vect{x} \in \mathbb{C}^{N \times 1}$ the transmitted signal from the BS antennas, which is written as $\vect{x}\triangleq\vect{V} \vect{s} + \vect{z}$, where $\vect{V} \in \mathbb{C}^{N \times N_d}$ is a linear precoding matrix and $\vect{s} \sim \mathcal{CN}(\mathbf{0},\mathbf{I}_{N_d})$ is the legitimate information symbol vector comprised by $N_d \leq \min\{M,N\}$ independent data streams, which is assumed independent from the AN vector $\vect{z} \in \mathbb{C}^{N \times 1}$ having the covariance matrix $\vect{Z}\triangleq\mathbb{E}\{\vect{z}\vect{z}^H\}$. In this paper, $N_d$ is considered an optimization variable, which will be jointly designed with the rest of the free parameters for the RIS-empowered legitimate link.
The baseband received signal vectors $\vect{y}_{\rm RX}\in\mathbb{C}^{M \times 1}$ and $\vect{y}_{\rm E}\in\mathbb{C}^{K\times 1}$ at the RX and Eve antenna elements, respectively, can be mathematically expressed as
\begin{align}
    \vect{y}_{\rm RX} &= \left(\vect{H} + \vect{H}_2 \vect{\Phi} \vect{H}_1\right) \left( \vect{V} \vect{s} +  \vect{z} \right) + \vect{n}_{\rm RX}, \label{eq:y_RX} \\
    \vect{y}_{\rm E} &= \left(\vect{H}_{\rm E} + \vect{G}_{\rm E} \vect{\Phi} \vect{H}_1 + \vect{G}_2 \vect{\Psi} \vect{G}_1 \right) \left( \vect{V} \vect{s} +  \vect{z} \right) + \vect{n}_{\rm E}, \label{eq:y_E}
\end{align}
where $\vect{\Phi}\triangleq\diag\{\vect{\phi}\}\in\mathbb{C}^{L \times L}$ and $\vect{\Psi}\triangleq\diag\{\vect{\psi}\}\in\mathbb{C}^{\Lambda \times \Lambda}$ with $\vect{\psi}\triangleq[e^{\jmath \xi_1}\,\,e^{\jmath \xi_2}\,\,\cdots\,\,e^{\jmath \xi_\Lambda} ]^T\in\mathbb{C}^{\Lambda\times 1}$ denoting the phase configuration of the malicious RIS in which $\xi_k$, with $k = 1,2,\ldots,\Lambda$, represents the phase shifting value at the $k$-th RIS unit element. In the latter two expressions, $\vect{n}_{\rm RX}\sim \mathcal{CN}(\vect{0}_{M\times1},\sigma^2 \vect{I}_M)$ and $\vect{n}_{\rm E}\sim \mathcal{CN}(\vect{0}_{K\times1},\sigma^2 \vect{I}_K)$ stand for the Additive White Gaussian Noise (AWGN) vectors, which model the thermal noises at the receivers. 

We assume the linear digital combining matrices $\vect{U} \in \mathbb{C}^{M \times N_d}$ and $\vect{W} \in \mathbb{C}^{K \times N_d}$ at the RX and Eve, respectively, which will be designed later on. This assumption considers that Eve knows the value $N_d$ of the independent data streams used in the legitimate link, which actually serves as an upper bound for Eve's achievable rate. Using the latter signal models, the achievable rates in bps/Hz at the legitimate and eavesdropping links are given by
\begin{align}
    \mathcal{R}_{\rm RX} &\triangleq \log_2 \left| \vect{I}_{N_d} \!+\! \vect{U}^H \tilde{\vect{H}} \vect{V}\vect{V}^H \tilde{\vect{H}}^H \vect{U} \left( \vect{U}^H \left( \sigma^2 \vect{I}_M \!+\! \tilde{\vect{H}} \vect{Z} \tilde{\vect{H}}^H  \right) \vect{U} \right)^{-1} \right|, \label{eq:R_RX} \\
    \mathcal{R}_{\rm E} &\triangleq \log_2 \left| \vect{I}_{N_d} \!+\! \vect{W}^H \tilde{\vect{H}}_{\rm E} \vect{V}\vect{V}^H \tilde{\vect{H}}_{\rm E}^H \vect{W} \left( \vect{W}^H \left( \sigma^2 \vect{I}_K \!+\! \tilde{\vect{H}}_{\rm E} \vect{Z} \tilde{\vect{H}}_{\rm E}^H  \right) \vect{W} \right)^{-1} \right|, \label{eq:R_E}
\end{align}
where $\tilde{\vect{H}}\triangleq\vect{H} + \vect{H}_2 \vect{\Phi} \vect{H}_1$ and $\tilde{\vect{H}}_{\rm E}\triangleq\vect{H}_{\rm E} + \vect{G}_{\rm E} \vect{\Phi} \vect{H}_1 + \vect{G}_2 \vect{\Psi} \vect{G}_1 $. The secrecy rate is then obtained as $\mathcal{R}_{\rm s}\triangleq\max\{0,\mathcal{R}_{\rm RX}-\mathcal{R}_{\rm E}\}$ \cite{Oggier_2011}.

\subsection{Design of the Eavesdropping Parameters $\vect{W}$ and $\vect{\psi}$}\label{E_design} 
We assume that Eve is unaware of the fact that BS transmits the AN vector $\vect{z}$, and jointly designs its combining matrix $\vect{W}$ and the malicious RIS's reflection vector $\vect{\psi}$, profiting from the perfect availability of $\vect{G}_2$ and the partial knowledge of $\vect{G}_1$. In particular, Eve leverages on the LOS placement of the malicious RIS relative to the BS to compose the LOS component of $\vect{G}_1$ as $\hat{\vect{G}}_1 \triangleq \vect{g}_1 \vect{g}_2^H$, where $\vect{g}_1 \in \mathbb{C}^{\Lambda \times 1}$ and $\vect{g}_2 \in \mathbb{C}^{N \times 1}$ represent the array response vectors at the BS and RIS, respectively.
To this end, it possesses the cascaded channel $\bar{\vect{H}}_{\rm E}\triangleq\vect{G}_2 \vect{\Psi} \hat{\vect{G}}_1$ and derives its baseband received signal $\bar{\vect{y}}_{\rm E}\in\mathbb{C}^{K\times 1}$ via the following expression\footnote{Note that \eqref{eq:y_bar} differs from the actual received signal expression \eqref{eq:y_E} that includes the AN and the impact of the legitimate RIS-empowered link. This happens because Eve is assumed to be unaware of the existence of the AN and legitimate RIS. Note that, to mitigate AN, high computational power would be needed from the eavesdropper side \cite{Niu_2022}, hence, Eve neglects it.}: 
\begin{align}\label{eq:y_bar}
\bar{\vect{y}}_{\rm E}\triangleq \bar{\vect{H}}_{\rm E} \vect{V} \vect{s} + \vect{n}_{\rm E}.
\end{align}
Note that Eve is unaware of the BS precoding vector $\vect{V}$ appearing in \eqref{eq:y_bar}, hence, we assume it considers an omni-directional precoder such that $\trace(\vect{V}^H \vect{V}) = P$, with $P$ being the BS transmit power. In addition, Eve is unaware of the existence of the legitimate RIS, and thus it neglects the reception of an explicit reflected signal via the channel $\vect{G}_{\rm E}$. Following the latter considerations, the eavesdropping system formulates the following joint design optimization problem:
\begin{align*}
    \mathcal{OP}_{\rm E}: \,\max_{{\vect{W}, \, \vect{\psi}}} \, & \quad \bar{\mathcal{R}}_{\rm E} \quad
    \text{s.t.}  \quad \norm{\vect{W}}_F^2 \leq 1,\,\,\lvert \psi_{k} \rvert = 1  \, \, \forall k = 1,2,\dots,\Lambda,
\end{align*}
where the rate $\bar{\mathcal{R}}_{\rm E}$ is given, using the definition $\mathbf{C}\triangleq\vect{W}^H \vect{W}$, by:
\begin{equation} \label{eq:bar_R_E}
\bar{\mathcal{R}}_{\rm E} \triangleq \log_2 \left| \vect{I}_{N_d} + \sigma^{-2} \vect{W}^H\bar{\vect{H}}_{\rm E}  \vect{V} \vect{V}^H \bar{\vect{H}}_{\rm E}^H \vect{W} \mathbf{C}^{-1} \right|,
\end{equation}
and the constraint for $\vect{W}$ restricts any undesired amplification of the thermal noise. To solve the latter problem, we adopt AO, similar to the approach that will be described in the following section for the optimization of the design parameters for the legitimate RIS-empowered link. In particular, the linear combiner matrix $\vect{W}$ is solved similarly to the linear combiner $\vect{U}$ of the legitimate design problem, as will be presented in Section \ref{Sec:Optimize_U}. Then, for $\vect{W}$ considered fixed and by replacing $\bar{\vect{H}}_E$, it can be shown that the rate expression in \eqref{eq:bar_R_E} reduces to:
\begin{equation*}
	\bar{\mathcal{R}}_{\rm E} \triangleq \log_2 \left( 1 + \tilde{\sigma}^{-2} \vect{g}_1^H \vect{\Psi}^H \vect{G}_2^H \vect{W} \mathbf{C}^{-1} \vect{W}^H \vect{G}_2 \vect{\Psi} \vect{g}_1 \right),
\end{equation*}
where $\tilde{\sigma}^{-2} \triangleq \sigma^{-2}\vect{g}_2^H \vect{V} \vect{V}^H \vect{g}_2$. Next, by defining $\hat{\vect{D}} \triangleq \diag\{\vect{g}_1\}^H \vect{G}_2^H \vect{W} \mathbf{C}^{-1} \vect{W}^H \vect{G}_2 \diag\{\vect{g}_1\}$, $\bar{\mathcal{R}}_{\rm E}$ can be rewritten as $\bar{\mathcal{R}}_{\rm E} = \log_2 (1 + \tilde{\sigma}^{-2} \vect{\psi}^H \hat{\vect{D}} \vect{\psi})$, whose solution can be attained by Riemanian MO, as will be presented in Section \ref{Sec:Optimize_phi}, after taking into account that the Euclidean gradient of $h(\vect{\psi}) \triangleq - \bar{\mathcal{R}}_{\rm E}(\vect{\psi})$ is given as $\nabla_{\vect{\psi}}h = -\frac{1}{\ln 2}\frac{2 \tilde{\sigma}^{-2} \hat{\vect{D}} \vect{\psi}}{1 + \tilde{\sigma}^{-2} \vect{\psi}^H \hat{\vect{D}} \vect{\psi}}$.

\section{Proposed RIS-Empowered MIMO Secrecy Design} \label{Sec:Prob_Form}
According to our system model, the BS lacks knowledge about the existence of any malicious RIS. Hence, its believed baseband received signal $\hat{\vect{y}}_{\rm E}\in\mathbb{C}^{K \times 1}$ at Eve is expressed as\footnote{Note that \eqref{eq:y_hat} differs from the actual received signal expression \eqref{eq:y_E}, which includes the signal received at Eve via the malicious RIS; recall that the legitimate system is unaware of its existence.}:
\begin{equation}\label{eq:y_hat}
\hat{\vect{y}}_{\rm E} \triangleq \left( \vect{H}_{\rm E} + \vect{G}_{\rm E} \vect{\Phi} \vect{H}_1 \right) \left( \vect{V} \vect{s} + \vect{z} \right) + \vect{n}_{\rm E}.
\end{equation}
Using the latter expression and assuming capacity-achieving combining at Eve (which serves as Eve's upper bound performance), the BS formulates Eve's instantaneous achievable rate, using the definition $\hat{\vect{H}}_{\rm E} \triangleq \vect{H}_{\rm E} + \vect{G}_{\rm E} \vect{\Phi} \vect{H}_1$, as the following function of $\vect{V}$ and $\vect{Z}$:
\begin{equation} \label{eq:R_E_bs}
 \hat{\mathcal{R}}_{\rm E,inst} \triangleq \log_2 \left\lvert \vect{I}_K + \hat{\vect{H}}_{\rm E} \vect{V} \vect{V}^H \hat{\vect{H}}_{\rm E}^H \left( \sigma^2 \vect{I}_K + \hat{\vect{H}}_{\rm E} \vect{Z} \hat{\vect{H}}_{\rm E}^H \right)^{-1} \right\rvert.
\end{equation}

In this paper, we consider the following secrecy rate maximization problem for the joint design of the legitimate BS linear precoding matrix $\vect{V}$ and the number of data streams $N_d$, the AN covariance matrix $\vect{Z}$, the linear combiner $\vect{U}$ at the legitimate RX, and the phase configuration vector $\vect{\phi}$ of the legitimate RIS:
\begin{align*}
    \mathcal{OP}_{\rm L}: \,\, \max_{N_d, \vect{U}, \vect{V}, \vect{Z} \succeq \vect{0},  \vect{\phi}} & \quad \hat{\mathcal{R}}_{\rm s} \triangleq \mathcal{R}_{\rm RX} - \mathbb{E}_{\vect{H}_{\rm E},\vect{G}_{\rm E}}\left[ \hat{\mathcal{R}}_{\rm E,inst} \right] \\
    \text{s.t.} & \quad \trace(\vect{V}^H \vect{V}) + \trace(\vect{Z}) \leq P,\,\,\lvert \phi_{\ell} \rvert = 1 \,\, \forall \ell = 1,2,\dots,L,\\
    & \quad \norm{\vect{U}}_F^2 \leq 1,\, 1 \leq N_d \leq \min\{M,N\}, 
\end{align*}
where $P$ denotes the total transmit power budget and the constraint for $\vect{U}$ excludes any undesired amplification of the reception thermal noise. In this problem formulation, $\mathbb{E}_{\vect{H}_{\rm E},\vect{G}_{\rm E}}[\cdot]$ represents the joint expectation over the channels $\vect{H}_{\rm E}$ and $\vect{G}_{\rm E}$, indicating the assumed statistical CSI knowledge that the BS possesses for the eavesdropping system (see Section~\ref{Sec:Threat_Model}). Since the objective function in $\mathcal{OP}_{\rm L}$ is difficult to handle, we next derive an upper bound for Eve's rate expression.
\begin{Prop} \label{Prop:Prop_Ergodic_R_E}
The ergodic achievable rate expression for Eve is upper bounded by:
\begin{equation}\label{eq:R_E_bs_ub}
	\mathbb{E}_{\vect{H}_{\rm E},\vect{G}_{\rm E}}\left[ \hat{\mathcal{R}}_{\rm E,inst} \right] \leq \hat{\mathcal{R}}_{\rm E}^{\rm ub} \triangleq \log_2 \left\lvert \vect{I}_N + \sigma^{-2} \vect{Q} \bar{\vect{X}} \right\rvert - \log_2 \left\lvert \vect{I}_N + \sigma^{-2} \vect{Q} \vect{Z} \right\rvert,
\end{equation}
where $\bar{\vect{X}} \triangleq \vect{V} \vect{V}^H + \vect{Z}$ and $\vect{Q} \triangleq \mathbb{E} \left[ \hat{\vect{H}}_{\rm E}^H \hat{\vect{H}}_{\rm E} \right] = \vect{Q}_{\vect{H}_{\rm E}} + \vect{H}_1^H \vect{\Phi}^H \vect{Q}_{\vect{G}_{\rm E}} \vect{\Phi} \vect{H}_1$, with $\vect{Q}_{\vect{H}_{\rm E}} \triangleq \mathbb{E}[\vect{H}_{\rm E}^H \vect{H}_{\rm E}] \in \mathbb{C}^{N \times N}$ and $\vect{Q}_{\vect{G}_{\rm E}} \triangleq \mathbb{E}[\vect{G}_{\rm E}^H \vect{G}_{\rm E}] \in \mathbb{C}^{L \times L}$.
\end{Prop}
\begin{IEEEproof}
	See Appendix~\ref{appx:Prop_Ergodic_R_E}.
\end{IEEEproof}

The resulting design problem for the legitimate system, after substituting the bound in \eqref{eq:R_E_bs_ub} into $\mathcal{OP}_{\rm L}$'s objective, is non-convex and can be solved via AO. To this end, we perform exhaustive search over all $\min\{M,N\}$ possible values for $N_d$ to find the one maximizing the objective function. In particular, for each feasible $N_d$ value, we transform it into the more tractable form of \cite[Lemma 4.1]{Shi_2015}, and perform AO over the involved variables, as described in the sequel.

\begin{Lem} \label{Lemma_AO_AN}
Suppose that $\vect{M} \in \mathbb{C}^{n \times n}$ with $\vect{M}\succeq\vect{0}$ is defined as:
\begin{equation} \label{eq:MSE_matrix}
\vect{M} \triangleq \left(\vect{A}^H \vect{B}\vect{C} - \vect{I}_n\right) \left(\vect{A}^H \vect{B}\vect{C} - \vect{I}_n\right)^H + \vect{A}^H\vect{R}\vect{A},
\end{equation}
where $\vect{A} \in \mathbb{C}^{m \times n}$, $\vect{B} \in \mathbb{C}^{m \times n}$, $\vect{C} \in \mathbb{C}^{n \times n}$, and $\vect{R} \in \mathbb{C}^{m \times m}$ with $\vect{R}\succ\vect{0}$. Let also the scalar function%
\footnote{The statement of Lemma \ref{Lemma_AO_AN} includes the natural logarithm $\log(\cdot)$, hence, the rates are expressed in nats/sec/Hz. The rate formulas in \eqref{eq:R_RX} and \eqref{eq:R_E}, which are expressed in bits/sec/Hz, can be converted to nats/sec/Hz via a multiplication by the factor $\log(2)$. Note, however, that the solution of $\mathcal{OP}_{\rm L}$ does not depend on this scaling factor.}  
$f(\vect{S},\vect{A})\triangleq \log \lvert\vect{S}\rvert - \trace(\vect{S}\vect{M}) + \trace(\vect{I}_N)$ with $\vect{S} \in \mathbb{C}^{n \times n}$. The following maximum values for $f(\vect{S},\vect{A})$ hold: 
\begin{align}
 &\log \lvert \vect{M}^{-1} \rvert = \max_{\vect{S} \succ \vect
 {0}} f(\vect{S},\vect{A}), \label{eq:lemma1-first-part} \\
& \log \left\lvert \vect{I}_n + (\vect{B} \vect{C})^H \vect{R}^{-1} \vect{B} \vect{C} \right\rvert = \max_{\vect{A}, \vect{S} \succ \vect{0}} f(\vect{S},\vect{A}), \label{eq:lemma1-second-part}
\end{align}
where the optimal values of \eqref{eq:lemma1-first-part} and \eqref{eq:lemma1-second-part} are obtained with the solution $\vect{S}_{\rm opt} = \vect{M}^{-1}$.
\end{Lem}

By introducing the auxiliary matrix variables $\vect{A}_1,\, \vect{S}_1 \in \mathbb{C}^{N_d \times N_d}$ and defining the following Mean Squared Error (MSE) matrix:
\begin{equation} \label{eq:MSE_matrix_RX}
\begin{aligned}
 \vect{M}_1 \triangleq \left(\vect{A}_1^H \vect{U}^H \tilde{\vect{H}} \vect{V} \!-\! \vect{I}_{N_d} \right) \left(\vect{A}_1^H \vect{U}^H \tilde{\vect{H}} \vect{V} \!-\! \vect{I}_{N_d} \right)^H + \vect{A}_1^H \left( \vect{U}^H \left(\sigma^2 \vect{I}_M + \tilde{\vect{H}} \vect{Z} \tilde{\vect{H}}^H\right) \vect{U} \right) \vect{A}_1,
\end{aligned}
\end{equation}
the achievable rate $\mathcal{R}_{\rm RX}$ in \eqref{eq:R_RX}, when expressed in nats/sec/Hz, can be equivalently rewritten as:
\begin{equation} \label{eq:equiv_R_RX}
 \mathcal{R}_{\rm RX} = N_d + \max_{\vect{A}_1, \vect{S}_1 \succ \vect{0}} \left(\log \lvert \vect{S}_1 \rvert - \trace(\vect{S}_1 \vect{M}_1)\right).
\end{equation}
To apply Lemma \ref{Lemma_AO_AN} for $\hat{\mathcal{R}}_{\rm E}^{\rm ub}$, again in nats/sec/Hz, we first capitalize on the fact that $\vect{Q} \succeq \vect{0}$ (by definition) and, using the notation $\vect{Q}^{H/2}\triangleq(\vect{Q}^{1/2})^H$, we re-express $\hat{\mathcal{R}}_{\rm E,1}^{\rm ub}$ and $\hat{\mathcal{R}}_{\rm E,2}^{\rm ub}$ as follows:
\begin{align}
 \hat{\mathcal{R}}_{\rm E,1}^{\rm ub} &= \log \left\lvert \vect{I}_N + \sigma^{-2} \vect{Q}^{H/2}\vect{Z} \vect{Q}^{1/2} \right\rvert, \label{eq:R_E_1}\\
 \hat{\mathcal{R}}_{\rm E,2}^{\rm ub} &= \log \left\lvert \vect{I}_N + \sigma^{-2} \vect{Q}^{H/2} \bar{\vect{X}} \vect{Q}^{1/2} \right\rvert. \label{eq:R_E_2}
\end{align}
Then, by expressing $\vect{Z}$ as $\vect{Z} \triangleq \tilde{\vect{Z}} \tilde{\vect{Z}}^H$ (e.g., via the eigenvalue decomposition) and introducing the auxiliary matrix variables $\vect{A}_2 \in \mathbb{C}^{K \times N}, \vect{S}_2 \in \mathbb{C}^{N \times N}$ and $\vect{S}_3 \in \mathbb{C}^{K \times K}$, the formula in \eqref{eq:R_E_1} can be rewritten as:
\begin{equation} \label{eq:R_E_part1}
 \hat{\mathcal{R}}_{{\rm E},1}^{\rm ub} = N + \max_{\vect{A}_2, \vect{S}_2 \succ \vect{0}} \left(\log \lvert\vect{S}_2 \rvert - \trace\left( \vect{S}_2 \vect{M}_2 \right)\right),
\end{equation}  
with $\vect{M}_2$ being the following MSE matrix:
\begin{equation}
 \vect{M}_2 \triangleq ( \vect{A}_2^H \vect{Q}^{H/2} \tilde{\vect{Z}} - \vect{I}_N )( \vect{A}_2^H \vect{Q}^{H/2} \tilde{\vect{Z}} - \vect{I}_N)^H + \sigma^2 \vect{A}_2^H \vect{A}_2.
\end{equation}
Similarly, \eqref{eq:R_E_2} can be re-expressed as the optimization:
\begin{equation} \label{eq:R_E_part2}
 -\hat{\mathcal{R}}_{{\rm E},2}^{\rm ub} = N + \max_{\vect{S}_3 \succ \vect{0}} \left(\log \lvert \vect{S}_3 \rvert -\trace\left( \vect{S}_3 \vect{M}_3 \right)\right),
\end{equation}
where $\vect{M}_3 \triangleq \vect{I}_N + \sigma^{-2} \vect{Q}^{H/2} \bar{\vect{X}} \vect{Q}^{1/2} $. Putting all above together, the $\mathcal{OP}_{\rm L}$, excluding the optimization over the number of streams $N_d$, is recast to the following design problem:
\begin{align*}
\begin{split}
\mathcal{OP}_{{\rm L},\mathbb{X}}: \,\, \max_{\mathbb{X}} \quad & \bar{\mathcal{R}}_{\rm s}^{\rm lb}\triangleq\mathcal{R}_{\rm RX} + \hat{\mathcal{R}}_{{\rm E},1}^{\rm ub} - \hat{\mathcal{R}}_{{\rm E},2}^{\rm ub} \\
\text{s.t.} \quad & \trace(\vect{V} \vect{V}^H) + \trace(\tilde{\vect{Z}}\tilde{\vect{Z}}^H) \leq P,\,\,\norm{\vect{U}}_F^2 \leq 1,\,\,\lvert \phi_\ell \rvert = 1 \,\, \forall \ell = 1,2,\dots,L,
\end{split}
\end{align*}
where $\mathbb{X} \triangleq \{ \vect{A}_i, \vect{S}_j \succ \vect{0}, \vect{U}, \vect{V}, \tilde{\vect{Z}}, \vect{\phi} \}$ with $i \in \{1,2\}$ and $j \in \{1,2,3\}$. $\mathcal{OP}_{{\rm L},\mathbb{X}}$ is still non-convex, due to the coupled variables, as well as the unit-modulus constraints. However, it is easy to perceive that it is convex when treating each set of variables separately (except $\vect{\phi}$). Hence, it can be solved by a block coordinate descent approach, as will be next presented.

\subsection{$\mathcal{OP}_{\rm L}$'s Optimization with Respect to $\vect{A}_1$ and $\vect{A}_2$} \label{Sec:Optimize_A_i} 
After some algebraic manipulations with \eqref{eq:equiv_R_RX} and \eqref{eq:R_E_part1}, and then setting their first order derivatives with respect to the auxiliary variables $\vect{A}_1$ and $\vect{A}_2$, respectively, equal to zero, the optimal values for $\vect{A}_1$ and $\vect{A}_2$ become:
\begin{align}
 \vect{A}_{1, \rm opt} &= \left( \sigma^2 \vect{U}^H \vect{U} + \vect{U}^H \tilde{\vect{H}} \left( \vect{V} \vect{V}^H + \tilde{\vect{Z}} \tilde{\vect{Z}}^H \right) \tilde{\vect{H}}^H \vect{U} \right)^{-1} \vect{U}^H \tilde{\vect{H}} \vect{V}, \label{eq:optimal_A_1} \\
 \vect{A}_{2, \rm opt} &= \left( \sigma^2 \vect{I}_N + \vect{Q}^{H/2} \tilde{\vect{Z}} \tilde{\vect{Z}}^H \vect{Q}^{1/2} \right)^{-1} \vect{Q}^{H/2} \tilde{\vect{Z}}. \label{eq:optimal_A_2}
\end{align}

\subsection{$\mathcal{OP}_{\rm L}$'s Optimization with Respect to $\vect{S}_1$, $\vect{S}_2$, and $\vect{S}_3$} \label{Sec:Optimize_S_j} 
By substituting $\vect{A}_{1, \rm opt}$ and $\vect{A}_{2, \rm opt}$ into \eqref{eq:equiv_R_RX} and \eqref{eq:R_E_part1}, respectively, and invoking the matrix inversion lemma, the optimal expressions for the auxiliary variables $\vect{S}_1$ and $\vect{S}_2$ in $\mathcal{OP}_{{\rm L},\mathbb{X}}$ are obtained as follows:
\begin{align}
 \vect{S}_{1, \rm opt} &= \vect{I}_{N_d} + \vect{V}^H \tilde{\vect{H}}^H \vect{U} \left( \sigma^2 \vect{U}^H \vect{U} + \vect{U}^H \tilde{\vect{H}} \tilde{\vect{Z}} \tilde{\vect{Z}}^H \tilde{\vect{H}}^H \vect{U} \right)^{-1} \vect{U}^H \tilde{\vect{H}} \vect{V}, \label{eq:optimal_S_1} \\
 \vect{S}_{2, \rm opt} &= \vect{I}_N + \sigma^{-2} \tilde{\vect{Z}}^H \vect{Q} \tilde{\vect{Z}}. \label{eq:optimal_S_2}
\end{align}
Finally, by using Lemma \ref{Lemma_AO_AN}, the optimal $\vect{S}_3$ is obtained as $\vect{S}_{3, \rm opt} = \vect{M}_3^{-1}$.

\subsection{$\mathcal{OP}_{\rm L}$'s Optimization with Respect to $\vect{U}$} \label{Sec:Optimize_U} 
The optimization variable $\vect{U}$ referring to the RX combining matrix appears only in the $\mathcal{R}_{\rm RX}$ expression in \eqref{eq:equiv_R_RX}. It, hence, suffices to obtain its Lagrangian function and then equate its first-order derivative with zero. To this end, let $\kappa \geq 0$ be the Lagrange multiplier and the Lagrangian of $\vect{U}$ is given by:
\begin{equation} \label{eq:Lagrangian_U}
\mathcal{L}_{\mathcal{OP}_{L,\vect{U}}}(\vect{U},\kappa) = -\trace(\vect{S}_1 \vect{M}_1) - \kappa (\trace(\vect{U}^H \vect{U}) - 1).
\end{equation}
After replacing $\vect{M}_1$ in \eqref{eq:Lagrangian_U}, using the definitions $\vect{E} \triangleq \sigma^2 \vect{I}_M + \tilde{\vect{H}} \vect{V} \vect{V}^H \tilde{\vect{H}}^H + \tilde{\vect{H}} \tilde{\vect{Z}} \tilde{\vect{Z}}^H \tilde{\vect{H}}^H$, $\vect{F} \triangleq \vect{A}_1 \vect{S}_1 \vect{A}_1^H$, and $\vect{J} \triangleq \tilde{\vect{H}} \vect{V} \vect{S}_1 \vect{A}_1^H $, and treating the terms irrelevant to $\vect{U}$ as constants, the following linear system is deduced: $\vect{E} \vect{U} \vect{F} + \kappa \vect{U} = \vect{J}$.
The optimal RX combining matrix $\vect{U}$ solving the latter problem is then derived as:
\begin{equation} \label{eq:optimal_U}
 \vect{U}_{\rm opt}(\kappa) = \left( \operatorname{vec}(\vect{I}_{N_d})^T \otimes \vect{I}_M \right) (\vect{I}_{N_d} \otimes \operatorname{vec}(\vect{U}_\kappa^{\star}) ),
\end{equation}
where $\operatorname{vec}(\vect{U}_\kappa^{\star}) = \left( \vect{F}^T \otimes \vect{E} + \kappa \vect{I}_{M N_d} \right)^{\dagger} \operatorname{vec}(\vect{J})$.
As clearly observed from the latter expression and \eqref{eq:optimal_U}, $\vect{U}$ depends on $\kappa$. To ensure the complementary slackness condition \cite{Boyd_2004}:
\begin{equation}
 \kappa^{\star} \left( \trace\left(\vect{U}^H_{\rm opt}(\kappa^{\star}) \vect{U}_{\rm opt}(\kappa^{\star})\right)  - 1 \right) = 0,
\end{equation}
the optimal $\kappa$, denoted by $\kappa^{\star}$, can be computed using the following Corollary.

\begin{Cor} \label{Thm:OP_L_Find_kappa}
Let $\bar{\vect{Q}} \vect{\Xi} \bar{\vect{Q}}^H$ be the eigendecomposition of $\vect{F}^T \otimes \vect{E}$, i.e., $\vect{\Xi}$ is an $M N_d \times M N_d$ diagonal matrix, whose elements are the eigenvalues of $\vect{F}^T \otimes \vect{E}$, and $\bar{\vect{Q}} \in \mathbb{C}^{M N_d \times M N_d}$ contains their corresponding eigenvectors. The Lagrange multiplier $\kappa^{\star}$ can be obtained from the solution of the equation:
\begin{equation} \label{eq:optimum_kappa}
 \sum_{p = 1}^{M N_d} \frac{[\tilde{\vect{Q}}]_{p,p}}{([\vect{\Xi}]_{p,p} + \kappa)^2} = N_d^{-1},
\end{equation}
where $\tilde{\vect{Q}} \triangleq \bar{\vect{Q}}^H \operatorname{vec}(\vect{J}) \operatorname{vec}(\vect{J})^H \bar{\vect{Q}}$.
\end{Cor}
\begin{IEEEproof}
  See Appendix \ref{appx:Thm_OP_L_Find_kappa}.
\end{IEEEproof}

It can be easily observed that the left-hand side of \eqref{eq:optimum_kappa} is monotonically decreasing for $\kappa \geq 0$. Hence, $\kappa^{\star}$ can be obtained using an one-dimensional search, e.g., the bisection method. Once $\kappa^{\star}$ is computed, it can be replaced in the expression for $\operatorname{vec}(\vect{U}_\kappa^{\star})$ to get the optimal $\vect{U}_{\rm opt}(\kappa^{\star})$, as shown in \eqref{eq:optimal_U}.


\subsection{$\mathcal{OP}_{\rm L}$'s Optimization with Respect to $\{\vect{V}, \tilde{\vect{Z}} \}$} \label{Sec:Optimize_V_Z}  
For the optimization over the BS precoding matrix $\vect{V}$ and the AN covariance matrix $\tilde{\vect{Z}}$, it suffices to use the Lagrangian function of the reformulated objective $\bar{\mathcal{R}}_s^{\rm lb}$ in $\mathcal{OP}_{\rm L, \mathbb{X}}$ and set its first-order derivatives with respect to $\vect{V}$ and $\tilde{\vect{Z}}$, respectively, equal to zero, resulting in the expressions:
\begin{align}
 \vect{V}_{\rm opt}^{\lambda} &= (\lambda \vect{I}_N + \vect{R}_{\vect{V}_1} )^{-1} \vect{R}_{\vect{V}_2} \label{eq:OP_L_V_opt} \\
  \tilde{\vect{Z}}_{\rm opt}^{\lambda} &= (\lambda \vect{I}_N + \vect{R}_{\tilde{\vect{Z}}_1} )^{-1} \vect{R}_{\tilde{\vect{Z}}_2}, \label{eq:OP_L_Z_opt}
\end{align}
where $\lambda\geq0$ is the Lagrange multiplier associated with the transmit power constraint and
\begin{align}
 \vect{R}_{\vect{V}_1} &\triangleq \mathbf{K} + \sigma^{-2} \vect{Q}^{1/2} \vect{S}_3 \vect{Q}^{H/2}, 
 \vect{R}_{\vect{V}_2} \triangleq \tilde{\vect{H}}^H \vect{U} \vect{A}_1 \vect{S}_1, \\
\vect{R}_{\tilde{\vect{Z}}_1} &\triangleq \mathbf{K} + \vect{Q}^{1/2} \vect{A}_2 \vect{S}_2 \vect{A}_2^H \vect{Q}^{H/2} + \sigma^{-2} \vect{Q}^{1/2} \vect{S}_3 \vect{Q}^{H/2}, 
\vect{R}_{\tilde{\vect{Z}}_2} \triangleq \vect{Q}^{1/2} \vect{A}_2 \vect{S}_2,
\end{align}
with $\vect{K}\triangleq\tilde{\vect{H}}^H \vect{U} \vect{A}_1 \vect{S}_1 \vect{A}_1^H \vect{U}^H \tilde{\vect{H}}$.
Clearly, the optimum pair $( \vect{V}_{\rm opt}^{\lambda}, \tilde{\vect{Z}}_{\rm opt}^{\lambda})$ depends on $\lambda$. Similar to the optimization with respect to $\vect{U}$ using Corollary~\ref{Thm:OP_L_Find_kappa}, by defining the eigendecompositions $\vect{R}_{\vect{V}_1}\triangleq\vect{P}_{\vect{V}} \vect{\Lambda}_{\vect{V}} \vect{P}_{\vect{V}}^H$ and $\vect{R}_{\tilde{\vect{Z}}_1} \triangleq \vect{P}_{\tilde{\vect{Z}}} \vect{\Lambda}_{\tilde{\vect{Z}}} \vect{P}_{\tilde{\vect{Z}}}^H$, the optimum $\lambda$ can be obtained by solving the following equation via a bisection algorithm:
\begin{equation}
 \sum_{q = 1}^N \left( \frac{[\tilde{\vect{P}}_{\vect{V}}]_{q,q}}{\left( [\vect{\Lambda}_{\vect{V}}]_{q,q} + \lambda \right)^2} + \frac{[\tilde{\vect{P}}_{\tilde{\vect{Z}}}]_{q,q}}{\left( [\vect{\Lambda}_{\tilde{\vect{Z}}}]_{q,q} + \lambda \right)^2} \right) = P,
\end{equation}
where $\tilde{\vect{P}}_{\vect{V}} \triangleq \vect{P}_{\vect{V}}^H \vect{R}_{\vect{V}_2} \vect{R}_{\vect{V}_2}^H \vect{P}_{\vect{V}}$ and $\tilde{\vect{P}}_{\tilde{\vect{Z}}} \triangleq \vect{P}_{\tilde{\vect{Z}}}^H \vect{R}_{\tilde{\vect{Z}}_2} \vect{R}_{\tilde{\vect{Z}}_2}^H \vect{P}_{\tilde{\vect{Z}}}$.

\subsection{$\mathcal{OP}_{\rm L}$'s Optimization with Respect to $\vect{\phi}$} \label{Sec:Optimize_phi} 
The recasted version $\mathcal{OP}_{{\rm L},\mathbb{X}}$ of the design problem $\mathcal{OP}_{\rm L}$ when focusing only on the maximization with respect to the legitimate RIS phase configuration, i.e., $\mathcal{OP}_{{\rm L},\vect{\phi}}$, requires finding the vector $\vect{\phi}$ optimizing the terms $\mathcal{R}_{\rm RX}$, $f_{\rm E,1}(\vect{\phi}) \triangleq \hat{\mathcal{R}}_{{\rm E},1}^{\rm ub}$, and $f_{\rm E,2}(\vect{\phi}) \triangleq -\hat{\mathcal{R}}_{{\rm E},2}^{\rm ub}$ given by \eqref{eq:equiv_R_RX}, \eqref{eq:R_E_1}, and \eqref{eq:R_E_2}, respectively. It is noted that the latter two terms are more difficult to deal with, because they depend on $\vect{\phi}$ through the matrix $\vect{Q}^{1/2}$. To overcome this, we devise an MM-based approach which constructs a surrogate function for each of these two terms. In the following, we first provide a more convenient equivalent representation for $\mathcal{R}_{\rm RX}$, and then, present the proposed MM framework for designing the surrogate functions for $f_{\rm E,1}$ and $f_{\rm E,2}$.
\subsubsection{Equivalent Representation for $\mathcal{R}_{\rm RX}(\vect{\phi})$}
Starting with $\mathcal{R}_{\rm RX}(\vect{\phi}) = -\trace(\vect{S}_1 \vect{M}_1)$ with $\vect{M}_1$ being an MSE matrix, the following expression is deduced after some algebraic manipulations:
\begin{align}\label{eq:R_RX_phi}
	\mathcal{R}_{\rm RX}(\vect{\phi}) =& 
		-\trace( \vect{S}_1 \vect{A}_1^H \vect{U}^H \vect{H} \vect{V} \vect{V}^H \vect{H}_1^H \vect{\Phi}^H \vect{H}_2^H \vect{U} \vect{A}_1 )
		- \trace( \vect{S}_1 \vect{A}_1^H \vect{U}^H \vect{H}_2 \vect{\Phi} \vect{H}_1 \vect{V} \vect{V}^H \vect{H}^H \vect{U \vect{A}_1} )\nonumber\\
		&- \trace( \vect{S}_1 \vect{A}_1^H \vect{U}^H \vect{H}_2 \vect{\Phi} \vect{H}_1 \vect{V} \vect{V}^H \vect{H}_1^H \vect{\Phi}^H \vect{H}_2^H \vect{U} \vect{A}_1 ) + \trace( \vect{S}_1 \vect{V}^H \vect{H}_1^H \vect{\Phi}^H \vect{H}_2^H \vect{U} \vect{A}_1 )\nonumber\\
		&- \trace( \vect{S}_1 \vect{A}_1^H \vect{U}^H \vect{H} \vect{Z} \vect{H}_1^H \vect{\Phi}^H \vect{H}_2^H \vect{U} \vect{A}_1 ) - \trace( \vect{S}_1 \vect{A}_1^H \vect{U}^H \vect{H}_2 \vect{\Phi} \vect{H}_1 \vect{Z} \vect{H}^H \vect{U} \vect{A}_1 ) \\
		&- \trace( \vect{S}_1 \vect{A}_1^H \vect{U}^H \vect{H}_2 \vect{\Phi} \vect{H}_1 \vect{Z} \vect{H}_1^H \vect{\Phi}^H \vect{H}_2^H \vect{U} \vect{A}_1 ) + \trace( \vect{S}_1 \vect{A}_1^H \vect{U}^H \vect{H}_2 \vect{\Phi} \vect{H}_1 \vect{V} ) + c_1,\nonumber
\end{align}
where the constant $c_1$ is defined as:
\begin{align}
	c_1 \triangleq \begin{aligned}[t]
		&-\trace(\vect{S}_1 \vect{A}_1^H \vect{U}^H \vect{H} \vect{V} \vect{V}^H \vect{H}^H \vect{U} \vect{A}_1) + \trace(\vect{S}_1 \vect{A}_1^H \vect{U}^H \vect{H} \vect{V}) + \trace(\vect{S}_1 \vect{V}^H \vect{H}^H \vect{U \vect{A}_1})\\
		&- \sigma^2\trace(\vect{S}_1 \vect{A}_1^H \vect{U}^H \vect{U} \vect{A}_1) - \trace(\vect{S}_1 \vect{A}_1^H \vect{U}^H \vect{H} \vect{Z} \vect{H}^H \vect{U} \vect{A}_1).
	\end{aligned}
\end{align}
By using the cyclic property of the trace operator and the following matrix definitions:
\begin{equation*}
	\begin{aligned}[t]
		\vect{D}_{1} &\triangleq \vect{H}_1 \vect{V} \vect{V}^H \vect{H}^H \vect{U} \vect{A}_1 \vect{S}_1 \vect{A}_1^H \vect{U}^H \vect{H}_2, \\
		\vect{D}_{4} &\triangleq \vect{H}_1 \vect{Z} \vect{H}_1^H,\,\vect{D}_{5} \triangleq \vect{H}_1 \vect{V} \vect{S}_1 \vect{A}_1^H \vect{U}^H \vect{H}_2,\\
	\end{aligned}
\qquad 
\begin{aligned}[t] 
	\vect{D}_{2} &\triangleq \vect{H}_2^H \vect{U} \vect{A}_1 \vect{S}_1 \vect{A}_1^H \vect{U}^H \vect{H}_2,\,\vect{D}_{3} \triangleq \vect{H}_1 \vect{V} \vect{V}^H \vect{H}_1^H,\\
	\vect{D}_{6} &\triangleq \vect{H}_1 \vect{Z} \vect{H}^H \vect{U} \vect{A}_1 \vect{S}_1 \vect{A}_1^H \vect{U}^H \vect{H}_2,
\end{aligned} 
\end{equation*}
the expression \eqref{eq:R_RX_phi} can be compactly rewritten as:
\begin{equation} \label{eq:R_RX_phi_compact}
	\begin{aligned}
	\mathcal{R}_{\rm RX}(\vect{\phi}) = 
		&-\trace \left( \vect{\Phi}^H(\vect{D}_{1}^H + \vect{D}_{6}^H - \vect{D}_{5}^H) \right) - \trace \left( \vect{\Phi}(\vect{D}_{1} + \vect{D}_{6} - \vect{D}_{5}) \right) \\
		&- \trace \left( \vect{\Phi}^H \vect{D}_{2} \vect{\Phi}(\vect{D}_{3} + \vect{D}_{4}) \right) + c_1.		
	\end{aligned}
\end{equation}

\subsubsection{Surrogate Function for $f_{\rm E,1}(\vect{\phi})$}
To obtain a surrogate function for $f_{\rm E,1}(\vect{\phi})$, we first note that \eqref{eq:R_E_1} can be re-expressed, after defining $\tilde{\vect{Q}}_{\vect{H}_{\rm E}} \triangleq \vect{I}_N + \sigma^{-2} \tilde{\vect{Z}}^H \vect{Q}_{\vect{H}_{\rm E}} \tilde{\vect{Z}}$, as follows:
\begin{align*}
	f_{\rm E,1}(\vect{\phi}) &= -\log\left\lvert \left(\tilde{\vect{Q}}_{\vect{H}_{\rm E}} + \sigma^{-2} \tilde{\vect{Z}}^H \vect{H}_1^H \vect{\Phi}^H \vect{Q}_{\vect{G}_{\rm E}} \vect{\Phi} \vect{H}_1 \tilde{\vect{Z}} \right)^{-1} \right\rvert \\
	&= -\log\left\lvert \tilde{\vect{Q}}_{\vect{H}_{\rm E}}^{-1} - \tilde{\vect{Q}}_{\vect{H}_{\rm E}}^{-1} \vect{F}_{\vect{G}_{\rm E}} (\sigma^2 \vect{I}_L + \vect{F}_{\vect{G}_{\rm E}}^H \tilde{\vect{Q}}_{\vect{H}_{\rm E}}^{-1} \vect{F}_{\vect{G}_{\rm E}} )^{-1} \vect{F}_{\vect{G}_{\rm E}}^H \tilde{\vect{Q}}_{\vect{H}_{\rm E}}^{-1} \right\rvert,
\end{align*}
where $\vect{F}_{\vect{G}_{\rm E}} \triangleq \tilde{\vect{Z}}^H \vect{H}_1^H \vect{\Phi}^H \vect{Q}_{\vect{G}_{\rm E}}^{1/2}$ and the second equality holds by invoking the matrix inversion lemma. We proceed by introducing the following Lemma \cite{sun2016majorization}, which results from the first-order Taylor expansion of the scalar function $\log\lvert \cdot \rvert$.

\begin{Lem} \label{Lemma_MM_1}
	The function $\log \left\lvert \vect{Y} \right\rvert$ for any matrix $\vect{Y}$ can be upper-bounded as:
	\begin{equation} \label{eq:Lemma_MM_1}
		\log \left\lvert \vect{Y} \right\rvert \leq \log \lvert \tilde{\vect{Y}} \rvert + \trace(\tilde{\vect{Y}}^{-1}(\vect{Y} - \tilde{\vect{Y}})),
	\end{equation}
	where $\tilde{\vect{Y}}$ is a given point and the equality is achieved when $\vect{Y} = \tilde{\vect{Y}}$.
\end{Lem}

By introducing the matrix $\tilde{\vect{M}}_2(\vect{\phi}) \triangleq \tilde{\vect{Q}}_{\vect{H}_{\rm E}}^{-1} - \tilde{\vect{Q}}_{\vect{H}_{\rm E}}^{-1} \vect{F}_{\vect{G}_{\rm E}} (\sigma^2 \vect{I}_L + \vect{F}_{\vect{G}_{\rm E}}^H \tilde{\vect{Q}}_{\vect{H}_{\rm E}}^{-1} \vect{F}_{\vect{G}_{\rm E}} )^{-1} \vect{F}_{\vect{G}_{\rm E}}^H \tilde{\vect{Q}}_{\vect{H}_{\rm E}}^{-1}$, considering a feasible point $\tilde{\vect{\phi}}$ satisfying the unit-modulus constraints of $\mathcal{OP}_{\rm L}$, and applying Lemma \ref{Lemma_MM_1}, we obtain after some straightforward algebraic manipulations the lower bound $f_{\rm E,1}(\vect{\phi}) \geq g_{\rm E,1}(\vect{\phi}|\tilde{\vect{\phi}}) + h_1(\tilde{\vect{\phi}})$, which includes the expressions:
\begin{align}
	g_{\rm E,1}(\vect{\phi}|\tilde{\vect{\phi}}) &= \trace\left( \tilde{\vect{M}}_2^{-1}(\tilde{\vect{\phi}}) \left(\tilde{\vect{Q}}_{\vect{H}_{\rm E}}^{-1} \vect{F}_{\vect{G}_{\rm E}} \left(\sigma^2 \vect{I}_L + \vect{F}_{\vect{G}_{\rm E}}^H \tilde{\vect{Q}}_{\vect{H}_{\rm E}}^{-1} \vect{F}_{\vect{G}_{\rm E}} \right)^{-1} \vect{F}_{\vect{G}_{\rm E}}^H \tilde{\vect{Q}}_{\vect{H}_{\rm E}}^{-1}\right)\!\right), \label{eq:R_E_1_surrogate1} \\ 
	h_1(\tilde{\vect{\phi}}) &= -\log\left\lvert \tilde{\vect{M}}_2(\tilde{\vect{\phi}}) \right\rvert -\trace\left( \tilde{\vect{M}}_2^{-1}(\tilde{\vect{\phi}}) \left(\tilde{\vect{Q}}_{\vect{H}_{\rm E}}^{-1} \tilde{\vect{F}}_{\vect{G}_{\rm E}} \left(\sigma^2 \vect{I}_L + \tilde{\vect{F}}_{\vect{G}_{\rm E}}^H \tilde{\vect{Q}}_{\vect{H}_{\rm E}}^{-1} \tilde{\vect{F}}_{\vect{G}_{\rm E}} \right)^{-1} \tilde{\vect{F}}_{\vect{G}_{\rm E}}^H \tilde{\vect{Q}}_{\vect{H}_{\rm E}}^{-1}\right)\!\right), \nonumber\label{eq:R_E_1_surrogate1_constant}
\end{align}
with $\tilde{\vect{F}}_{\vect{G}_{\rm E}} \triangleq \tilde{\vect{Z}}^H \vect{H}_1^H \tilde{\vect{\Phi}}^H \vect{Q}_{\vect{G}_{\rm E}}^{1/2}$. However, the expression for $g_{\rm E,1}(\vect{\phi}|\tilde{\vect{\phi}})$ is difficult to handle since $\vect{\phi}$, which is encapsulated in $\vect{F}_{\vect{G}_{\rm E}}$, appears in the inverse matrix factor. We, next, approximate this function using the following Lemma.
\begin{Lem} \label{Lemma_MM_2}
	Let us define the scalar function $f(\vect{X},\vect{Y}) = \trace(\vect{A}\vect{X}\vect{Y}^{-1}\vect{X}^H)$ with $\vect{A} \in \mathbb{C}^{n\times n}$ being positive semi-definite, $\vect{Y} \in \mathbb{C}^{m \times m}$ is positive definite, and $\vect{X} \in \mathbb{C}^{n \times m}$. Then, the following inequality holds:
	\begin{equation}
		\begin{aligned}
			f(\vect{X},\vect{Y}) \geq& \trace\left(\vect{A}\tilde{\vect{X}}\tilde{\vect{Y}}^{-1}\tilde{\vect{X}}^H\right) - \trace\left(\vect{A}\tilde{\vect{X}}\tilde{\vect{Y}}^{-1}(\vect{Y} - \tilde{\vect{Y}})\tilde{\vect{Y}}^{-1}\tilde{\vect{X}}^H\right) \\ 
			&+ \trace\left( \vect{A}\tilde{\vect{X}}\tilde{\vect{Y}}^{-1}(\vect{X} - \tilde{\vect{X}})^H \right) + \trace\left( \vect{A}(\vect{X} - \tilde{\vect{X}})\tilde{\vect{Y}}^{-1}\tilde{\vect{X}}^H \right),
		\end{aligned}
	\end{equation}
	where $\tilde{\vect{X}}$ and $\tilde{\vect{Y}}$ share the same properties with $\vect{X}$ and $\vect{Y}$, respectively. The equality is achieved at the point $(\vect{X},\vect{Y}) = (\tilde{\vect{X}},\tilde{\vect{Y}})$.
\end{Lem}

\begin{IEEEproof}
	The function $f(\vect{X},\vect{Y}) = \trace\left(\vect{A}\vect{X}\vect{Y}^{-1}\vect{X}^H\right)$ for $\vect{A} \succeq \vect{0}$ and $\vect{Y} \succ \vect{0}$ is jointly convex with respect to $\vect{X}$ and $\vect{Y}$. Hence, it can be lower bounded by its linear expansion around the point $(\tilde{\vect{X}},\tilde{\vect{Y}})$.
\end{IEEEproof}
By applying Lemma \ref{Lemma_MM_2} using the substitutions $\vect{A} = \tilde{\vect{M}}_2^{-1}(\tilde{\vect{\phi}})$, $\vect{X} = \tilde{\vect{Q}}_{\vect{H}_{\rm E}}^{-1} \vect{F}_{\vect{G}_{\rm E}}$, and $\vect{Y} = \sigma^2 \vect{I}_L + \vect{F}_{\vect{G}_{\rm E}}^H \tilde{\vect{Q}}_{\vect{H}_{\rm E}}^{-1} \vect{F}_{\vect{G}_{\rm E}}$ to $g_{\rm E,1}(\vect{\phi}|\tilde{\vect{\phi}})$ in \eqref{eq:R_E_1_surrogate1}, we derive after some  algebraic operations the lower bound $g_{\rm E,1}(\vect{\phi}|\tilde{\vect{\phi}}) \geq \bar{g}_{\rm E,1}(\vect{\phi}|\tilde{\vect{\phi}}) + \bar{h}_1(\tilde{\vect{\phi}})$, where using the definition $\vect{J}_{\vect{G}_{\rm E}} \triangleq (\sigma^2 \vect{I}_L + \tilde{\vect{F}}_{\vect{G}_{\rm E}}^H \tilde{\vect{Q}}_{\vect{H}_{\rm E}}^{-1} \tilde{\vect{F}}_{\vect{G}_{\rm E}} )^{-1} \tilde{\vect{F}}_{\vect{G}_{\rm E}}^H \tilde{\vect{Q}}_{\vect{H}_{\rm E}}^{-1}$:
\begin{align}
	\begin{split} \label{eq:R_E_1_surrogate2}
		\bar{g}_{\rm E,1}(\vect{\phi}|\tilde{\vect{\phi}}) = {}& -\trace\left( \tilde{\vect{M}}_2^{-1}(\tilde{\vect{\phi}}) \vect{J}_{\vect{G}_{\rm E}} \vect{F}_{\vect{G}_{\rm E}}^H \tilde{\vect{Q}}_{\vect{H}_{\rm E}}^{-1} \vect{F}_{\vect{G}_{\rm E}} \vect{J}_{\vect{G}_{\rm E}}^H \right) \\
		& + \trace\left( \tilde{\vect{M}}_2^{-1}(\tilde{\vect{\phi}}) \vect{J}_{\vect{G}_{\rm E}} \vect{F}_{\vect{G}_{\rm E}}^H \tilde{\vect{Q}}_{\vect{H}_{\rm E}}^{-1} \right) + \trace\left( \tilde{\vect{M}}_2^{-1}(\tilde{\vect{\phi}}) \tilde{\vect{Q}}_{\vect{H}_{\rm E}}^{-1} \vect{F}_{\vect{G}_{\rm E}} \vect{J}_{\vect{G}_{\rm E}}^H \right),
	\end{split}\\
	\bar{h}_1(\tilde{\vect{\phi}}) ={}& \trace\left( \tilde{\vect{M}}_2^{-1}(\tilde{\vect{\phi}}) \vect{J}_{\vect{G}_{\rm E}} \tilde{\vect{F}}_{\vect{G}_{\rm E}}^H \tilde{\vect{Q}}_{\vect{H}_{\rm E}}^{-1} \tilde{\vect{F}}_{\vect{G}_{\rm E}} \vect{J}_{\vect{G}_{\rm E}}^H \right) - \trace\left( \tilde{\vect{M}}_2^{-1}(\tilde{\vect{\phi}}) \tilde{\vect{Q}}_{\vect{H}_{\rm E}}^{-1} \tilde{\vect{F}}_{\vect{G}_{\rm E}} \vect{J}_{\vect{G}_{\rm E}}^H \right). \label{eq:R_E_1_surrogate2_constant}
\end{align}

Putting all above together, we can conclude at the lower bound $f_{\rm E,1}(\vect{\phi}) \geq \bar{g}_{\rm E,1}(\vect{\phi}|\tilde{\vect{\phi}}) + \bar{h}_1(\tilde{\vect{\phi}}) + h_1(\tilde{\vect{\phi}})$. Expanding matrix $\vect{F}_{\vect{G}_{\rm E}}$ which depends on $\vect{\phi}$, and making use of trace's cyclic property, the following compact expression for $\bar{g}_{\rm E,1}(\vect{\phi}|\tilde{\vect{\phi}})$ is deduced:
\begin{align} \label{eq:R_E_1_pro_compact}
		\bar{g}_{\rm E,1}(\vect{\phi}|\tilde{\vect{\phi}}) = &-\trace\left( \vect{\Phi}^H \vect{Q}_{\vect{G}_{\rm E}}^{1/2} \vect{J}_{\vect{G}_{\rm E}}^H \tilde{\vect{M}}_2^{-1}(\tilde{\vect{\phi}}) \vect{J}_{\vect{G}_{\rm E}} \vect{Q}_{\vect{G}_{\rm E}}^{H/2} \vect{\Phi} \vect{H}_1 \tilde{\vect{Z}} \tilde{\vect{Q}}_{\vect{H}_{\rm E}}^{-1} \tilde{\vect{Z}}^H \vect{H}_1^H \right) \\
		&+ \trace\left( \vect{\Phi} \vect{H}_1 \tilde{\vect{Z}} \tilde{\vect{Q}}_{\vect{H}_{\rm E}}^{-1} \tilde{\vect{M}}_2^{-1}(\tilde{\vect{\phi}}) \vect{J}_{\vect{G}_{\rm E}} \vect{Q}_{\vect{G}_{\rm E}}^{H/2} \right) + \trace\left( \vect{\Phi}^H \vect{Q}_{\vect{G}_{\rm E}}^{1/2} \vect{J}_{\vect{G}_{\rm E}}^H \tilde{\vect{M}}_2^{-1}(\tilde{\vect{\phi}}) \tilde{\vect{Q}}_{\vect{H}_{\rm E}}^{-1} \tilde{\vect{Z}}^H \vect{H}_1^H \right).\nonumber
\end{align}
Finally, using the matrix definitions:
\begin{align*}
	\vect{D}_7 &\triangleq \vect{Q}_{\vect{G}_{\rm E}}^{1/2} \vect{J}_{\vect{G}_{\rm E}}^H \tilde{\vect{M}}_2^{-1}(\tilde{\vect{\phi}}) \vect{J}_{\vect{G}_{\rm E}} \vect{Q}_{\vect{G}_{\rm E}}^{H/2},\,\vect{D}_8 \triangleq \vect{H}_1 \tilde{\vect{Z}} \tilde{\vect{Q}}_{\vect{H}_{\rm E}}^{-1} \tilde{\vect{Z}}^H \vect{H}_1^H, \\
	\vect{D}_9 &\triangleq \vect{H}_1 \tilde{\vect{Z}} \tilde{\vect{Q}}_{\vect{H}_{\rm E}}^{-1} \tilde{\vect{M}}_2^{-1}(\tilde{\vect{\phi}}) \vect{J}_{\vect{G}_{\rm E}} \vect{Q}_{\vect{G}_{\rm E}}^{H/2},
\end{align*}
as well as $c_2(\tilde{\vect{\phi}}) \triangleq \bar{h}_1(\tilde{\vect{\phi}}) + h_1(\tilde{\vect{\phi}})$, expression \eqref{eq:R_E_1_pro_compact} can be re-written more compactly as:
\begin{equation} \label{eq:R_E_1_compact}
	\bar{g}_{\rm E,1}(\vect{\phi}|\tilde{\vect{\phi}}) = -\trace\left( \vect{\Phi}^H \vect{D}_7 \vect{\Phi} \vect{D}_8 \right) + \trace\left( \vect{\Phi}^H \vect{D}_9^H \right) + \trace\left( \vect{\Phi} \vect{D}_9 \right) + c_2(\tilde{\vect{\phi}}).
\end{equation}

\subsubsection{Surrogate Function for $f_{\rm E,2}(\vect{\phi})$}
To design an approximate function for $f_{\rm E,2}(\vect{\phi}) = -\hat{\mathcal{R}}_{{\rm E},2}^{\rm ub}(\vect{\phi})$, we first manipulate expression \eqref{eq:R_E_2} as:
\begin{align*}
	f_{\rm E,2}(\vect{\phi}) &= -\log \left\lvert \vect{I}_N + \sigma^{-2} \bar{\vect{X}}^{H/2} \vect{Q} \bar{\vect{X}}^{1/2} \right\rvert \\
	&= -\log \left\lvert \vect{I}_N + \sigma^{-2} \bar{\vect{X}}^{H/2} \left( \vect{Q}_{\rm \vect{H}_{\rm E}} + \vect{H}_1^H \vect{\Phi}^H \vect{Q}_{\vect{G}_{\rm E}} \vect{\Phi} \vect{H}_1 \right) \bar{\vect{X}}^{1/2} \right\rvert,
\end{align*}
where the first equality follows from the fact that $\bar{\vect{X}}\succeq \vect{0}$ (by definition) and the second one by expanding $\vect{Q}$. Then, by setting $\tilde{\vect{M}}_3(\tilde{\vect{\phi}}) = \vect{I}_N + \sigma^{-2} \bar{\vect{X}}^{H/2} \left( \vect{Q}_{\rm \vect{H}_{\rm E}} + \vect{H}_1^H \tilde{\vect{\Phi}}^H \vect{Q}_{\vect{G}_{\rm E}} \tilde{\vect{\Phi}} \vect{H}_1 \right) \bar{\vect{X}}^{1/2} $ and applying Lemma \ref{Lemma_MM_1}, we obtain the lower bound $f_{\rm E,2}(\vect{\phi}) \geq g_{\rm E,2}(\vect{\phi}|\tilde{\vect{\phi}}) + h_2(\tilde{\vect{\phi}})$ with
\begin{align} 
	g_{\rm E,2}(\vect{\phi}|\tilde{\vect{\phi}}) &= -\sigma^{-2} \trace\left(\vect{\Phi}^H \vect{Q}_{\vect{G}_{\rm E}} \vect{\Phi} \vect{H}_1 \bar{\vect{X}}^{1/2} \tilde{\vect{M}}_3^{-1}(\tilde{\vect{\phi}}) \bar{\vect{X}}^{H/2} \vect{H}_1^H \right), \label{eq:R_E_2_surrogate} \\ 
	h_2(\tilde{\vect{\phi}}) &= -\log\left\lvert \tilde{\vect{M}}_3(\tilde{\vect{\phi}}) \right\rvert + \sigma^{-2} \trace\left(\tilde{\vect{M}}_3^{-1}(\tilde{\vect{\phi}}) \bar{\vect{X}}^{H/2} \vect{H}_1^H \tilde{\vect{\Phi}}^H \vect{Q}_{\vect{G}_{\rm E}} \tilde{\vect{\Phi}} \vect{H}_1 \bar{\vect{X}}^{1/2}\right). \label{eq:R_E_2_constant}
\end{align}
We finally use the definitions $\vect{D}_{10} \triangleq \sigma^{-2}\vect{H}_1 \bar{\vect{X}}^{1/2} \tilde{\vect{M}}_3^{-1}(\tilde{\vect{\phi}}) \bar{\vect{X}}^{H/2} \vect{H}_1^H $ and $c_3(\tilde{\vect{\phi}}) \triangleq h_2(\tilde{\vect{\phi}})$ to derive the following expression:
\begin{equation} \label{eq:R_E_2_compact}
	\bar{g}_{\rm E,2}(\vect{\phi}|\tilde{\vect{\phi}}) = -\trace\left( \vect{\Phi}^H \vect{Q}_{\vect{G}_{\rm E}} \vect{\Phi} \vect{D}_{10} \right) + c_3(\tilde{\vect{\phi}}).
\end{equation}

\subsubsection{Lower Bound for $\hat{\mathcal{R}}_{\rm s}^{\rm lb}$ and Solution for $\vect{\phi}$}
Putting together \eqref{eq:R_RX_phi_compact}, \eqref{eq:R_E_1_compact}, and \eqref{eq:R_E_2_compact}, and invoking the matrix identities from \cite[Th. 1.11]{Zhang_2017}, $\mathcal{OP}_{{\rm L},\vect{\phi}}$'s objective function $\hat{\mathcal{R}}_{\rm s}^{\rm lb}(\vect{\phi})$ can be lower bounded as follows:
\begin{equation}\label{eq:Lower_Bound}
 \hat{\mathcal{R}}_{\rm s}^{\rm lb}(\vect{\phi}) \geq g(\vect{\phi}|\tilde{\vect{\phi}}) \triangleq \mathcal{R}_{\rm RX}(\vect{\phi}) + \bar{g}_{\rm E,1}(\vect{\phi}|\tilde{\vect{\phi}}) + \bar{g}_{\rm E,2}(\vect{\phi}|\tilde{\vect{\phi}}) \!=\! -\left(\vect{\phi}^H \vect{T} \vect{\phi} + 2 \Re\{\vect{\phi}^H \vect{v}^* \} + \sum_{i=1}^3c_i(\tilde{\vect{\phi}}) \right)\!,
\end{equation}
where we have used the matrix definitions:
\begin{align}
  \vect{T} &\triangleq \vect{D}_{2} \odot \left( \vect{D}_{3} + \vect{D}_{4} \right)^T + \vect{D}_{7} \odot \vect{D}_{8}^T + \vect{Q}_{\vect{G}_{\rm E}} \odot \vect{D}_{10}^T,\\
  \vect{v} &\triangleq \operatorname{vec}_{\rm d} \left( \vect{D}_{1} + \vect{D}_{6} - \vect{D}_{5} - \vect{D}_{9} \right). 
\end{align}
Optimizing with respect to $\vect{\phi}$ is still hard to tackle, even when using the bound $g(\vect{\phi}|\tilde{\vect{\phi}})$, due to the non-convexity of the unit-modulus constraint for all elements of $\vect{\phi}$. We thus adopt Riemannian MO \cite{Absil_2008} to solve it efficiently, by expressing the set of its constraints as the Cartesian product of $L$ complex circles, also known as a Riemannian submanifold in $\mathbb{C}^{L\times1}$. In particular, we denote each complex circle as $\mathcal{CC} \triangleq \{ \phi_{\ell} \in \mathbb{C}: \lvert \phi_{\ell} \rvert = 1 \}$, with $\ell=1,2,\dots,L$, and their product, that represents our problem's feasible set, as $\mathcal{CC}_{\rm M} = \mathcal{CC} \times \mathcal{CC} \times \dots \times \mathcal{CC}$. 
\begin{algorithm}[!t]
\begin{algorithmic}[1]
\caption{MO-Based Solution for optimizing with respect to $\vect{\phi}$}
\label{alg:OP_L_phi}
\State \textbf{Input:} $\vect{A}_i$ with $i\in\{1,2\}$, $\vect{S}_j$ with $j\in\{1,2,3\}$, $\vect{U}$, $\vect{V}$, $\vect{Z}$, $\epsilon>0$, $\rho > 0$, $\mu,\nu\in (0,1)$, and $\vect{\phi}_0$.
\State Compute $\vect{q}_0 = - \nabla_{\vect{\phi}}^{\rm R} g(\vect{\phi}|\vect{\phi}_0)$.
\For{$ n = 1, 2, \dots$}
    \State 
    \parbox[t]{\dimexpr\linewidth-\algorithmicindent}{%
    The Armijo-Goldstein backtracking line search: Find the smallest integer $\omega \geq 0$ such that $g\left(\operatorname{unit}(\vect{\phi}_{n-1} + \rho\nu^{\omega}\vect{q}_{n-1})|\vect{\phi}_{n-1} \right) - g(\vect{\phi}_{n-1}|\vect{\phi}_{n-1}) \leq \mu \rho \nu^{\omega} \Re\{ (\nabla_{\vect{\phi}}^{\rm R} g(\vect{\phi}|\vect{\phi}_{n-1}))^H \vect{q}_{n-1} \}$.
    }
    \State Compute step size $\tau_{n-1} = \rho \nu^{\omega}$.    
    \State Compute $\hat{\vect{\phi}}_n = \vect{\phi}_{n-1} + \tau_{n-1} \vect{q}_{n-1}$ and $\vect{\phi}_n = \operatorname{unit}(\hat{\vect{\phi}}_n)$.
    \State Compute $\vect{q}_n$ and the Polak-Ribière constant $\zeta_{n-1}$ according to \eqref{eq:CG_direction_phi} and \eqref{eq:Polak_Ribiere_zeta_n}, respectively.
    
    \If $\norm{\nabla_{\vect{\phi}}^{\rm R}g(\vect{\phi}|\vect{\phi}_n)}^2 \leq \epsilon$
        \State $\vect{\phi}^{\star} = \vect{\phi}_n$ and \textbf{break};
    \EndIf
\EndFor
\State \textbf{Output:} $\vect{\phi}^{\star}$.
\end{algorithmic}
\end{algorithm}

The algorithmic steps of the proposed MO approach for solving with respect to $\vect{\phi}$ are summarized in Algorithm~\ref{alg:OP_L_phi}. At each $n$-th iterative step of this algorithm, the step size $\tau_{n-1}$ is first computed, and then, $\vect{\phi}_n$ is obtained by the retraction operator $\operatorname{unit}(\cdot)$ \cite[Sec. 4.1]{Absil_2008}. Next, the conjugate gradient descent direction $\vect{q}_n$ is derived as:
\begin{equation} \label{eq:CG_direction_phi}
 \vect{q}_n = - \nabla_{\vect{\phi}}^{\rm R} g(\vect{\phi}|\vect{\phi}_n) + \zeta_{n-1} \mathbf{T}_{n-1 \rightarrow n}(\vect{q}_{n-1}),
\end{equation}
where $\vect{\phi}_n$ represents the legitimate RIS phase configuration vector at the $n$-th step, and $\mathbf{T}_{n-1 \rightarrow n}$ is defined for any vector $\vect{r}$ as follows:
\begin{equation}
 \mathbf{T}_{n-1 \rightarrow n}(\vect{r}) \triangleq \vect{r} - \Re \{ \vect{r} \odot (\vect{\phi}_n^T)^H \} \odot \vect{\phi}_n.
\end{equation}
Specifically, $\mathbf{T}_{n-1 \rightarrow n}(\vect{q}_{n-1})$ represents the operation which is used in order to map the vector $\vect{q}_{n-1}$ from the tangent space\footnote{The tangent space is defined as $\mathcal{T}_{\vect{\phi}} \triangleq \{ \vect{u} \in \mathbb{C}^{L\times1} : \operatorname{vec}_{\rm d}(\vect{u} \vect{\phi}^H) = \vect{0} \}$.} $\mathcal{T}_{\vect{\phi}_{n-1}}$ to the tangent space $\mathcal{T}_{\vect{\phi}_{n}}$ of the $\mathcal{CC}_{\rm M}$. In addition, $\zeta_{n-1}$ is the Polak-Ribière parameter, which is used to achieve faster convergence and is given by 
\begin{equation} \label{eq:Polak_Ribiere_zeta_n}
\zeta_{n-1} = \frac{\Re{\left\{\left( \nabla_{\vect{\phi}}^{\rm R} g(\vect{\phi}|\vect{\phi}_n) \right)^H \left( \nabla_{\vect{\phi}}^{\rm R} g(\vect{\phi}|\vect{\phi}_n) - \mathbf{T}_{n-1 \rightarrow n}(\nabla_{\vect{\phi}}^{\rm R} g(\vect{\phi}|\vect{\phi}_{n-1})) \right)  \right\}}}{\norm{\nabla_{\vect{\phi}}^{\rm R} g(\vect{\phi}|\vect{\phi}_{n-1})}^2}.
\end{equation}
The Riemannian gradient in \eqref{eq:CG_direction_phi} and \eqref{eq:Polak_Ribiere_zeta_n}, which is the orthogonal projection of the Euclidean gradient to the tangent space $\mathcal{T}_{\vect{\phi}}$ of the $\mathcal{CC}_{\rm M}$, is given by
\begin{equation} \label{eq:Riemannian_Grad}
 \nabla_{\vect{\phi}}^{\rm R} g(\vect{\phi}|\tilde{\vect{\phi}}) = \nabla_{\vect{\phi}} g(\vect{\phi}|\tilde{\vect{\phi}}) - \Re\{\nabla_{\vect{\phi}} g(\vect{\phi}|\tilde{\vect{\phi}}) \odot (\vect{\phi}^T)^H \} \odot \vect{\phi}.
\end{equation}
The latter expression indicates that, to compute the Riemannian gradient, it suffices to calculate the Euclidean gradient of $\mathcal{OP}_{{\rm L},\vect{\phi}}$'s lower bound $g(\vect{\phi}|\tilde{\vect{\phi}})$ on $\hat{\mathcal{R}}_{\rm s}^{\rm lb}(\vect{\phi})$, which is given by $\nabla_{\vect{\phi}} g(\vect{\phi}|\tilde{\vect{\phi}}) = 2(\vect{T} \vect{\phi} + \vect{v}^*)$.

\subsection{Proposed Secrecy Design Algorithm} \label{Sec:OP_L_Overall_Algorithm} 
\begin{algorithm}[!t]
\begin{algorithmic}[1]
\caption{Proposed Secrecy Design Solving $\mathcal{OP}_{\rm L}$}
\label{alg:OP_L_Overall_Algorithm}
\State \textbf{Input:} $p=0$, $\epsilon > 0$, as well as feasible $\vect{V}^{(0)}$, $\vect{Z}^{(0)}$, $\vect{\phi}^{(0)}$, and $\hat{\mathcal{R}}_{\rm s}^{(0)}$ as defined in $\mathcal{OP}_{\rm L}$.
\For{$ m = 1,2,\ldots,\min\{M,N\}$} 
    \For{$ p = 1,2,\dots$}
    \State Compute $\tilde{\vect{H}} = \vect{H}_2 \diag{\{\vect{\phi}^{(p-1)}_m\}} \vect{H}_1$.
    \State Compute $\vect{A}_i^{(p)}$ with $i \in \{1,2\}$ using \eqref{eq:optimal_A_1} and \eqref{eq:optimal_A_2}.
		\State Compute $\vect{S}_{1}^{(p)}$ using \eqref{eq:optimal_S_1}, $\vect{S}_{2}^{(p)}$ using \eqref{eq:optimal_S_2}, and $\vect{S}_3^{(p)} = \left(\vect{M}_3^{(p)}\right)^{-1}$.
		\State Compute $\vect{U}^{(p)}_m$ using the expression for $\operatorname{vec}(\vect{U}_\kappa^{\star})$, \eqref{eq:optimal_U}, and a bisection method.	
    \State Compute $\vect{V}^{(p)}_m$ and $\tilde{\vect{Z}}^{(p)}$ according to \eqref{eq:OP_L_V_opt}, \eqref{eq:OP_L_Z_opt}, and a bisection method.
		\State Set $\vect{Z}^{(p)}_m = \tilde{\vect{Z}}^{(p)} \left(\tilde{\vect{Z}}^{{(p)}}\right)^H$.
    \State Obtain $\vect{\phi}^{(p)}_m$ using Algorithm \ref{alg:OP_L_phi}.
    \If $\left\lvert\left(\hat{\mathcal{R}}_{\rm s}^{(p)} - \hat{\mathcal{R}}_{\rm s}^{(p-1)}\right)/\hat{\mathcal{R}}_{\rm s}^{(p)}\right\rvert \leq \epsilon$, \textbf{break}; 
    \EndIf
    \EndFor
\State Compute $\hat{\mathcal{R}}_{\rm s}^{(p)}$ for $N_d=m$ streams using $\vect{U}_m^{(p)}$, $\vect{V}_m^{(p)}$, $\vect{Z}_m^{(p)}$, and $\vect{\phi}_m^{(p)}$.
\EndFor
\State Choose $N_d = m^{\star}$ with $m^{\star}$ yielding the maximum rate.
\State \textbf{Output:} $\vect{U}_{m^{\star}}^{(p)}$, $\vect{V}_{m^{\star}}^{(p)}$, $\vect{Z}_{m^{\star}}^{(p)}$, and $\vect{\phi}_{m^{\star}}^{(p)}$.
\end{algorithmic}
\end{algorithm}

The algorithmic steps of the proposed block coordinate descent approach for solving $\mathcal{OP}_{\rm L}$ are summarized in Algorithm~\ref{alg:OP_L_Overall_Algorithm}. The convergence properties of the inner iterative loop of this algorithm (i.e., for each number $m\leq\min\{M,N\}$ of independent data streams) are characterized by the following theorem.
\begin{Thm} \label{Thm:OP_L_KKT_Convergence}
The Algorithm \ref{alg:OP_L_Overall_Algorithm} for solving $\mathcal{OP}_{\rm L}$ has a non-decreasing trend and its output point $(\vect{V}^{(I)}_m, \vect{Z}^{(I)}_m, \vect{U}^{(I)}_m, \vect{\phi}^{(I)}_m)$, with $I$ being the number of iterations for the inner loop convergence, satisfies the Karush–Kuhn–Tucker (KKT) conditions of the problem $\forall$$m\in\{1,2,\ldots,\min\{M,N\}\}$.
\end{Thm}
\begin{IEEEproof}
The proof is delegated in Appendix~\ref{appx:Thm_OP_L_KKT_Convergence}.
\end{IEEEproof}

The computational complexity of Algorithm~\ref{alg:OP_L_Overall_Algorithm} is analyzed via inspection of its algorithmic steps, as follows. In step $4$, the product of three matrices with the middle one being diagonal requires $\mathcal{O}(ML(N + 1))$ computations. The calculation of the auxiliary matrix variables in steps $5$ and $6$ is dominated by the inversion of the involved matrices, resulting in $\mathcal{O}(\max\{N_d^3,K^3,N^3\})$ computational complexity, which deduces to $\mathcal{O}(N^3)$ since $N>K>N_d$ (the number of BS antennas is usually larger than those at Eve and the RX terminals). In step $7$, the worst case complexity is $\mathcal{O}(M^3N_d^3)$, due to the matrix inversion operation and the eigenvalues' computation. For calculating $\vect{V}$ and $\vect{Z}$ in steps $8$ and $9$, the required computational cost is $\mathcal{O}(2N^3)$. Note that the computation of the optimal Lagrange multipliers via the bisection method is negligible. In step $10$, the proposed MO in Algorithm~1 is used that requires $\mathcal{O}(I_{\rm MO} L^{1.5})$ complexity \cite{Shewchuk_1994} with $I_{\rm MO}$ denoting this algorithm's convergence iteration number. This cost mainly comes from the computation of the Euclidean gradient. Hence, the total complexity of Algorithm~\ref{alg:OP_L_Overall_Algorithm} for solving the proposed secrecy design $\mathcal{OP}_{\rm L}$ is
\begin{equation}
 C_{\mathcal{OP}_{\rm L}} = \mathcal{O}( I \min\{M,N\} \max\{ ML(N + 1), 2 N^3, M^3N_d^3, I_{\rm MO} L^{1.5} \} ),
\end{equation}
where the factor $\min\{M,N\}$ refers to the exhaustive search for finding the optimum number of the transmitted data streams (i.e., the outer loop of the algorithm).

\section{Numerical Results and Discussion} \label{Sec:Numerical} 
\begin{figure}[!t]
\centering
\includegraphics[scale=0.9]{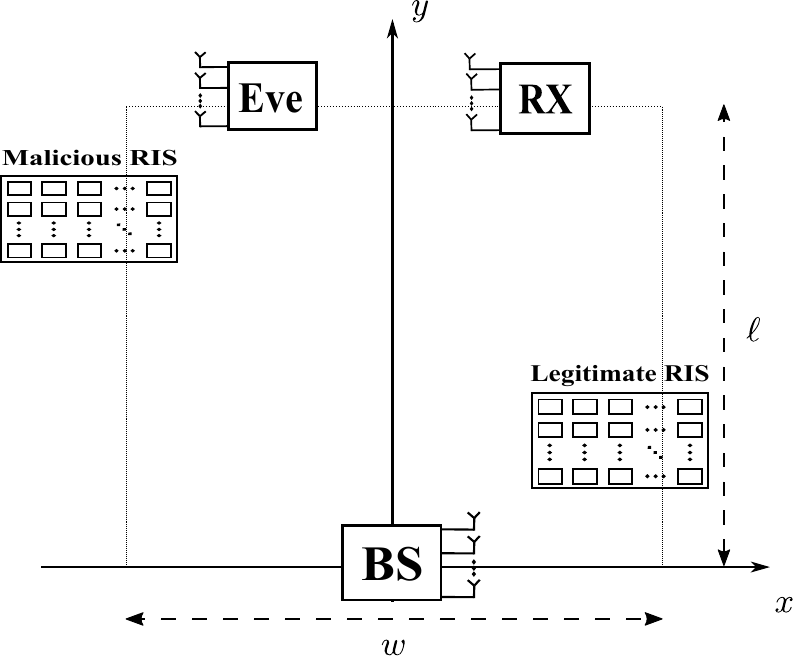}
\caption{\small{The $xy$-plane of the simulated RIS-empowered MIMO PLS system in $3$D. Each node's coordinates $(x,y,h)$ include the distances $x$ and $y$ along the horizontal and vertical axes, respectively, and the value $h$ in the $z$-axis (i.e., node's height). Each node is placed on the perimeter of a rectangle with width $w$ and length $\ell$. }}
\label{fig:Sims_Model}
\end{figure}
In this section, we investigate the secrecy rate performance of the proposed PLS scheme, by numerically evaluating the actual achievable rates of the legitimate and eavesdropping links using expressions \eqref{eq:R_RX} for $\mathcal{R}_{\rm RX}$ and \eqref{eq:R_E} for $\mathcal{R}_{\rm E}$, respectively, as well as $\mathcal{R}_{\rm s}$ providing the secrecy rate. We have adopted the proposed scheme in Section~\ref{E_design} for the RX combining and the RIS reflective beamforming of the eavesdropping system. For the legitimate system, we have used the proposed PLS scheme in Section~\ref{Sec:Prob_Form}, which encompasses BS precoding and AN, RX combining, and legitimate RIS reflective beamforming, as well as a special version of it for the case where a legitimate RIS is not available. For this special version, we have solved a similar problem to $\mathcal{OP}_{\rm L}$ via Lemma \ref{Lemma_AO_AN} and block coordinate descent, by removing the links involving the legitimate RIS and the optimization over its relevant variable $\vect{\phi}$. As a benchmark scheme, we have considered the PLS design of \cite{Conf} (termed as ``Perfect ECSI''), according to which the BS is assumed to have perfect knowledge of the matrices $\vect{H}_{\rm E}$ and $\vect{G}_{\rm E}$, and Eve perfectly knows $\vect{G}_1$ via respective cooperation with the legitimate BS.

In our simulations, all nodes were considered positioned on a $3$-Dimensional ($3$D) coordinate system, whose $xy$-plane is illustrated in Fig$.$~\ref{fig:Sims_Model}. As depicted, a rectangle of width $w$ and length $\ell$, with $\ell > w$, is used for each node's placement. The coordinates of each node are given by the triad $(x,y,h)$, where $x$ denotes the coordinate of the node on the $x$-axis, $y$ is the coordinate on the $y$-axis, and $h$ represents the node's height (i.e., each point's value on the $z$-axis, which is not shown in Fig$.$~\ref{fig:Sims_Model}). We have assumed that the BS, RX, Eve, the malicious RIS ${\rm RIS}_M$, and the legitimate RIS ${\rm RIS}_L$ are located at $(0,0,10)\,m$, $(\frac{w}{4},\ell,1.5)\,m$, $(-\frac{w}{4},\ell,1.5)\,m$, $(-\frac{w}{2},\frac{7\ell}{8},5)\,m$, and $(\frac{w}{2},y_{\rm RIS_L},5)\,m$, respectively, with $y_{\rm RIS_L} > 0$ denoting the position of the legitimate RIS on the $y$-axis. We have also considered distance dependent pathloss between any two nodes $i$ and $j$ with distance $d_{ij}$ (where $i$ and $j$ take values from the string set $\{{\rm BS},{\rm RX},{\rm E},{\rm RIS}_M,{\rm RIS}_L\}$), which was modeled as ${\rm PL}_{ij} = {\rm PL}_0 (d_{ij}/d_0)^{-\varepsilon_{ij}}$, where ${\rm PL}_0 = - 30$ dB denotes the pathloss at the reference distance $1$ $m$ and $\varepsilon_{ij}$ is the pathloss exponent. All wireless channels were modeled as flat Rician faded (with Rician factor denoted by $\kappa$) according to the following expression:
\begin{equation} \label{eq:actual_channels}
	\vect{H}_{ij} = \sqrt{\rm PL_{ij}} \left( \sqrt{\frac{\kappa}{\kappa+1}}\vect{H}_{ij}^{\rm LOS} + \sqrt{\frac{1}{\kappa+1}}\vect{H}_{ij}^{\rm NLOS} \right),
\end{equation}
where $\vect{H}_{ij}^{\rm LOS}$ and $\vect{H}_{ij}^{\rm NLOS}$ represent the LOS and Non-LOS (NLOS) channel components, respectively. The former was modeled via the response of a Uniform Planar Array (UPA). In particular, the UPA response, when having $A \triangleq A_v \times A_h$ antenna elements (where $A_v$ and $A_h$ denote the number of vertical and horizontal antennas, respectively), is given by $\vect{\alpha}(\theta,\varphi,A_v,A_h) \triangleq \vect{a}_{A_v}(\theta) \otimes \vect{a}_{A_h}(\theta,\varphi)$, with $\theta$ and $\varphi$ denoting the inclination and azimuth angles of arrival (departure), respectively, and $\vect{a}_{A_v}(\cdot)$ and $\vect{a}_{A_h}(\cdot,\cdot)$ being the $A_v$- and $A_h$-element array steering vectors\cite[eqs. (2) and (3)]{STAR_RIS_2022}, respectively. Thus, $\vect{H}_{ij}^{\rm LOS} = \vect{\alpha}(\theta_j,\varphi_j,A_{v_j},A_{h_j}) \vect{\alpha}^H(\theta_i,\varphi_i,A_{v_i},A_{h_i})$, following the convention that $i$ refers to the transmitter and $j$ to the receiver. The NLOS component was modeled via the Rayleigh distribution, specifically $[\vect{H}_{ij}^{\rm NLOS}]_{m,n} \sim \mathcal{CN}(0,1)$ $\forall m,n$. It is noted that the model in \eqref{eq:actual_channels} was used to compute the actual achievable rates using \eqref{eq:R_RX} and \eqref{eq:R_E}.

To model the statistical CSI availability for the channels $\vect{H}_{\rm E}$ and $\vect{G}_{\rm E}$ involving Eve, we assume that both are defined according to the model \cite{Wang_2021}: $\vect{\Upsilon} \sim \mathcal{CN}(\vect{M},\vect{\Sigma}\otimes\vect{\Xi})$, or equivalently $\vect{\Upsilon} = \vect{M} + \vect{\Sigma}^{1/2} \vect{\Upsilon}_w \vect{\Xi}^{(1/2)T}$, with $\vect{\Upsilon}_w \in \mathbb{C}^{m\times n}$ being a complex Gaussian matrix with independent and identically distributed zero-mean and unit-variance entries, and $\vect{\Sigma} \in \mathbb{C}^{m\times m}$ and $\vect{\Xi} \in \mathbb{C}^{n\times n}$ being Hermitian matrices;  clearly, $\vect{\Upsilon}$ will be a $m\times n$ complex Gaussian matrix.  
For the matrices $\vect{\Sigma}$ and $\vect{\Xi}$, we use the Kronecker separable correlation model according to \cite[eqs. (10) and (11)]{Wang_2021}, which is suitable for UPAs. 
To this end, by letting $\vect{M}_{\vect{H}_{\rm E}} = \sqrt{\operatorname{PL}_{\rm BS,\rm E}} \sqrt{\frac{\kappa}{\kappa + 1}} \vect{H}_{\rm E}^{\rm LOS}$, the distribution of the channel $\vect{H}_{\rm E}$ was modeled as (similarly for $\vect{G}_{\rm E}$):
\begin{equation}
	\vect{H}_{\rm E} \sim \mathcal{CN}\left(\vect{M}_{\vect{H}_{\rm E}},\vect{\Sigma}_{\vect{H}_{\rm E}} \otimes \frac{\operatorname{PL}_{\rm BS,\rm E}}{\kappa + 1} \vect{\Xi}_{\vect{H}_{\rm E}}^T \right).
\end{equation}
It can be easily shown using \cite{gupta2018matrix} that the $\vect{Q}_{\vect{H}_{\rm E}}$ (similarly $\vect{Q}_{\vect{G}_{\rm E}}$), required in \eqref{eq:Lower_Bound}, is given by:
\begin{equation}
	\vect{Q}_{\vect{H}_{\rm E}} = \vect{M}_{\vect{H}_{\rm E}}^H \vect{M}_{\vect{H}_{\rm E}} + \frac{\operatorname{PL}_{\rm BS,\rm E}}{\kappa + 1} \trace\left( \vect{\Sigma}_{\vect{H}_{\rm E}} \right)  \vect{\Xi}_{\vect{H}_{\rm E}}^T.
\end{equation}

In the performance results that follow, two different setups were investigated: \textit{i}) Setup (a) that considers the placement of the legitimate RIS close to the BS with $y_{\rm RIS_L} = \ell/8$, $N=8$ and $K = 4$; and \textit{ii}) Setup (b) where the legitimate RIS is close to the RX and Eve with $y_{\rm RIS_L} = 7\ell/8$, $N=16$, and $K = 8$. In both setups, we have set the width of the rectangle as $w = 15\,m$ and $\ell = 90\,m$, the number of $\rm RX$ antennas as $M = 4$, the pathloss exponents as $\varepsilon_{ij} = 5$ for $i = \rm BS$ and $j \in \{\rm RX, E\}$ and $\varepsilon_{ij} = 2$ for all other links, the Rician factor as $\kappa = 13.2$ dB, the noise variance at the RX and Eve as $\sigma^2 = -105$ dBm, and the convergence threshold for both Algorithms~1 and~2 as $\epsilon = 10^{-6}$. We have used $500$ independent Monte Carlo realizations for all performance curves in the following figures.

\subsection{Absence of a Legitimate RIS and Existence of one Malicious RIS} \label{Sec:Numerical_AN_only} 
\begin{figure}[!t]
    \centering
    \begin{subfigure}[h]{0.48\textwidth}
        \includegraphics[width=\textwidth]{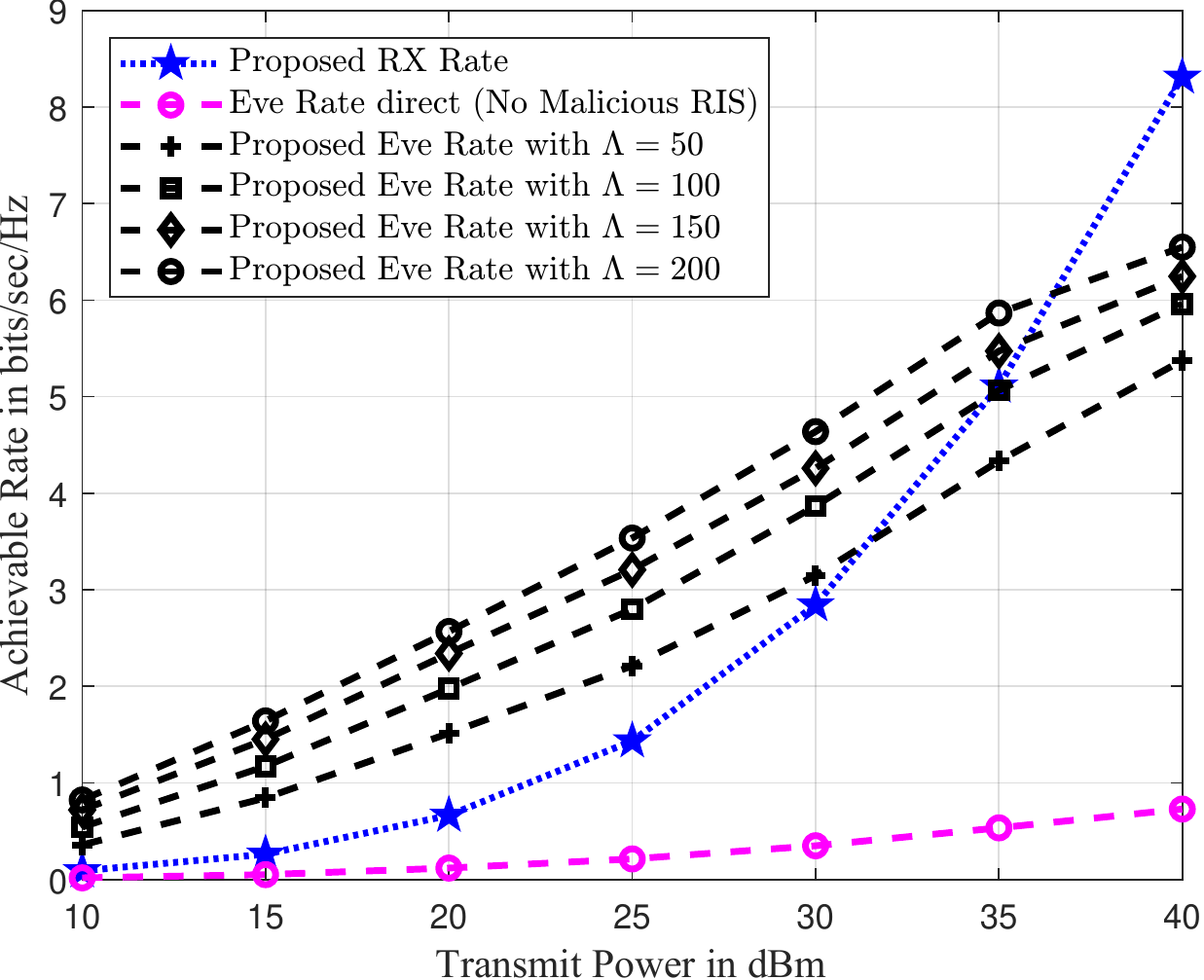}
        \caption{\small{PLS system Setup (a) without a legitimate RIS.}}
        \label{fig:AN_ONLY_A}
    \end{subfigure}
    ~ 
			\,
    \begin{subfigure}[h]{0.48\textwidth}
        \includegraphics[width=\textwidth]{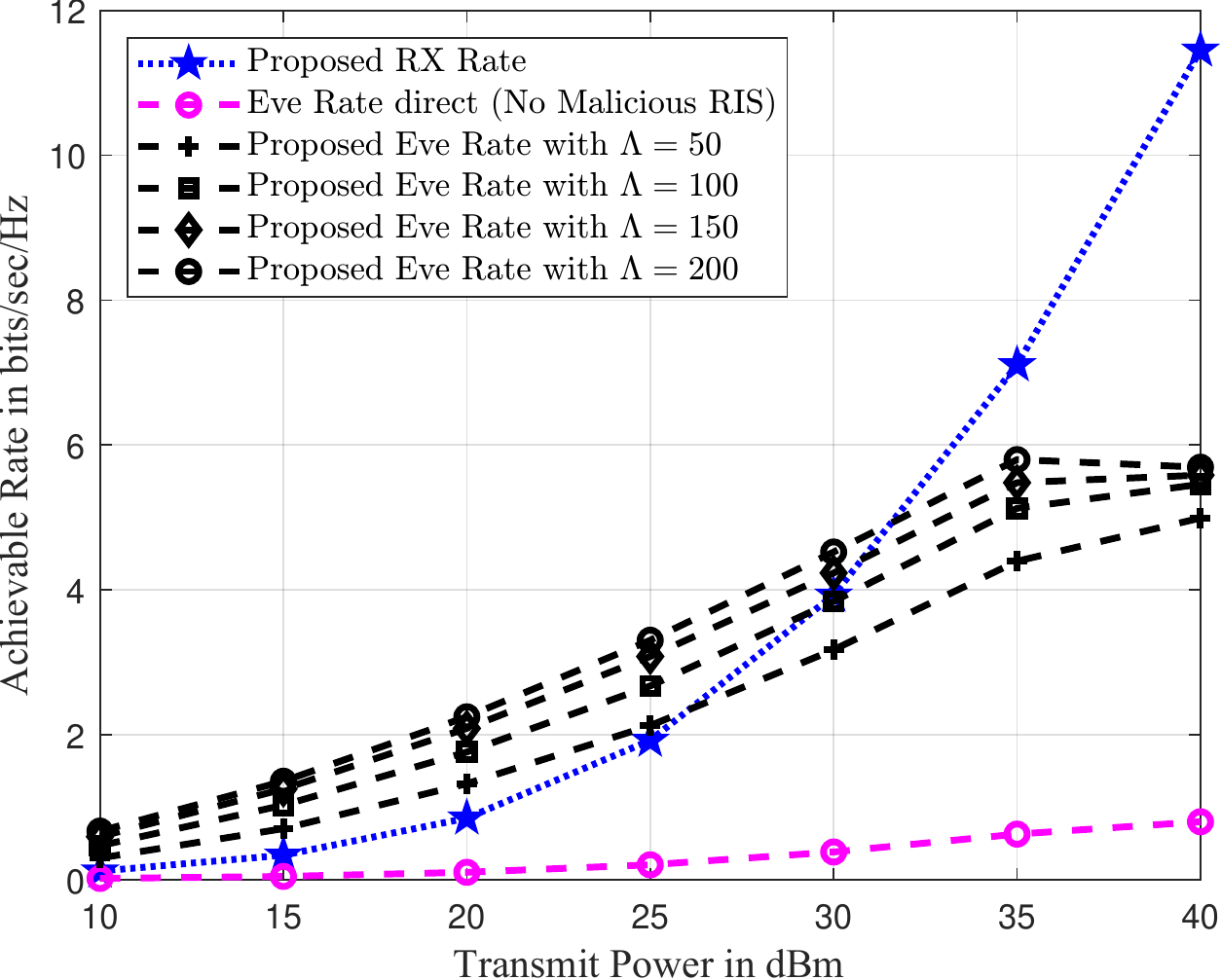}
        \caption{\small{PLS system Setup (b) without a legitimate RIS.}}
        \label{fig:AN_ONLY_B}
    \end{subfigure}
    \caption{\small{Achievable rates in bits/sec/Hz at the legitimate RX and the eavesdropper Eve versus the transmit power $P$ in dBm for both simulated PLS system setups, without a legitimate RIS and different numbers $\Lambda$ for the unit elements at the malicious RIS. The legitimate system intends to safeguard confidential communication with only BS precoding and AN, as well as RX combining.}}
\label{fig:AN_ONLY_Both}
\end{figure} 
We first consider the case where the legitimate system does not include an RIS and targets at securing confidential transmissions with only BS precoding, AN, and RX combining, via the solution of the aforementioned special case of $\mathcal{OP}_{\rm L}$. In Fig$.$~\ref{fig:AN_ONLY_Both}, we depict the achievable rates in bps/Hz at both RX and Eve as functions of the BS transmit power $P$ in dBm for both considered setups and various values $\Lambda$ for the elements of the malicious RIS. It can be seen for both scenarios that all rates increase with increasing $P$, while for $P>35$ dBm, the achievable rates at Eve saturate for any $\Lambda$ value. This saturation, combined with the larger RX rate values than those at Eve for all considered $\Lambda$'s, indicates the desirable role of AN in safeguarding legitimate communications. It is also shown for both setups that Eve's rate increases with increasing $\Lambda$. More specifically, for $P=[10,35]$ dBm, the rates at Eve for $\Lambda=\{100,150,200\}$ in Setup (a) are greater than, or equal to, the achievable rate at RX, whereas in Setup (b) this trend happens only for $P=[10,30]$ dBm and $\Lambda=\{150,200\}$. The latter behavior implies that the secrecy rate is zero, verifying the prominent role of the malicious RIS, via the presented eavesdropping scheme in Section~\ref{E_design}, in boosting Eve's capability to correctly decode legitimate information. As also depicted in the figure, when this malicious RIS is absent, the rates at Eve are very low, hence, eavesdropping becomes impossible. Clearly, when $\Lambda$ is large and $P$ moderate to high, the proposed PLS scheme falls short in safeguarding the legitimate link. The latter behavior happens due to the fact that BS is unaware of the presence of the malicious RIS (which can have $\Lambda \gg N$ unit elements \cite{HMIMO}), and only possesses statistical knowledge of the matrix $\vect{H}_{\rm E}$ for the design of the parameters of the legitimate link.

\begin{figure}[!t]
	\centering
	\begin{subfigure}[h]{0.48\textwidth}
		\includegraphics[width=\textwidth]{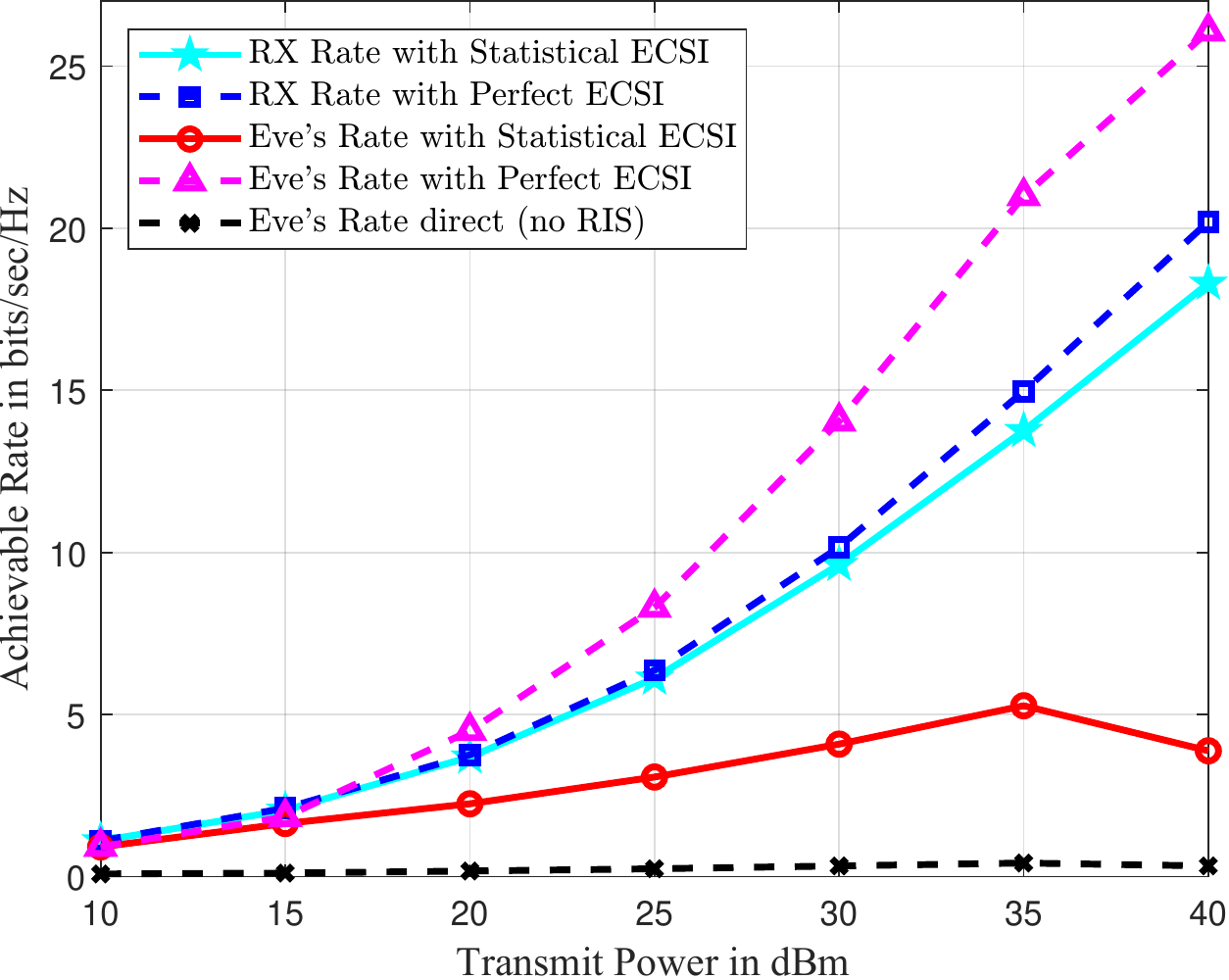}
		\caption{\small{Achievable rates in bits/sec/Hz at the RX and Eve versus the transmit power $P$ in dBm for the simulated PLS Setup (a), considering both the statistical and perfect ECSI knowledge cases with $L = 20$ and $\Lambda = 100$.}}
		\label{fig:Rates_Full_A}
	\end{subfigure}
	~ 
	\,
	\begin{subfigure}[h]{0.48\textwidth}
		\includegraphics[width=\textwidth]{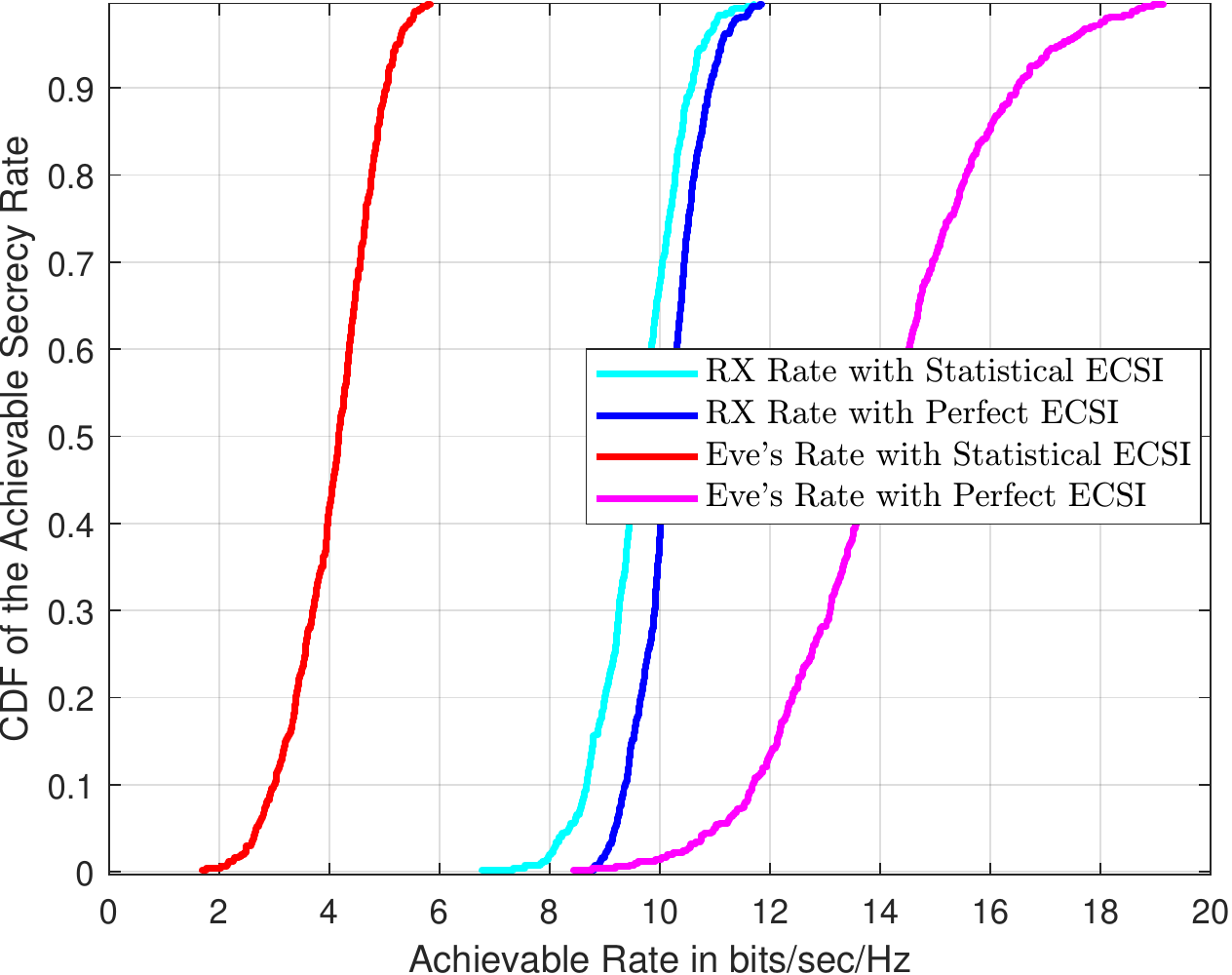}
		\caption{\small{CDF of the achievable rates at the RX and Eve for both the statistical and perfect ECSI knowledge cases, considering the transmit power $P=30$ dBm and the Setup (a) with $L = 20$ and $\Lambda = 100$.}}
		\label{fig:CDF_Setup_A_B}
	\end{subfigure}
	\caption{\small{Performance comparison between the proposed PLS scheme, relying on statistical ECSI knowledge at the BS, and the Perfect ECSI scheme of \cite{Conf}. In the former scheme, the proposed threat model is used, according to which Eve knows partially the channel between the BS and the malicious RIS, whereas for the latter scheme, Eve knows perfectly all channels required for its rate computation, except the one resulting from the legitimate RIS.}}
	\label{fig:Comparisons_Stat_Perf_ECSI}
\end{figure} 

\subsection{Co-Existence of one Legitimate and one Malicious RISs}\label{Sec:Numerical_Full_Problem} 
Considering the existence of an $L$-element legitimate RIS, we compare, in Fig.~\ref{fig:Comparisons_Stat_Perf_ECSI}, the achievable rates between the proposed PLS scheme, using the presented threat model in Section~\ref{Sec:Threat_Model} and relying on statistical ECSI knowledge at the BS, and the Perfect ECSI scheme of \cite{Conf}. As depicted in Fig.~\ref{fig:Rates_Full_A} for the Perfect ECSI case and the considered values of the elements of the RISs (where $\Lambda=5L$), the achievable rate for Eve dominates over the one for RX for all $P$ values, implying zero secrecy rates.
However, when Eve knows partially $\vect{G}_1$, its achievable rate via the proposed eavesdropping design in Section~\ref{E_design} gets severely degraded, while the RX rate with statistical ECSI and the proposed scheme in Section~\ref{Sec:Prob_Form} performs close to that with perfect ECSI knowledge. This behavior results in increasing positive secrecy rates with increasing $P$, verifying the safeguarding capability of the proposed PLS scheme for the case of statistical ECSI availability and non-cooperating eavesdropping. We further investigate this trend in Fig.~\ref{fig:CDF_Setup_A_B}, where we compare the empirical Cumulative Distribution Functions (CDF) of the achievable rate with both schemes, considering the transmit power level $P=30$ dBm and the Setup (a) with $L = 20$ and $\Lambda = 100$. 
As illustrated, the distribution of the rate concerning the legitimate link is very similar for both ECSI knowledge cases, indicating the effectiveness of our proposed PLS functionality relying solely on reduced CSI. It is also shown that there exists a large gap between the distributions of Eve's rate for the two ECSI cases, which witnesses the the safequarding capability of our PLS scheme for the more practical case where Eve does not cooperate with the BS to be to able to estimate the $\vect{G}_1$ channel.
\begin{figure}[!t]
\centering
\begin{subfigure}[h]{0.48\textwidth}
	\includegraphics[width=\textwidth]{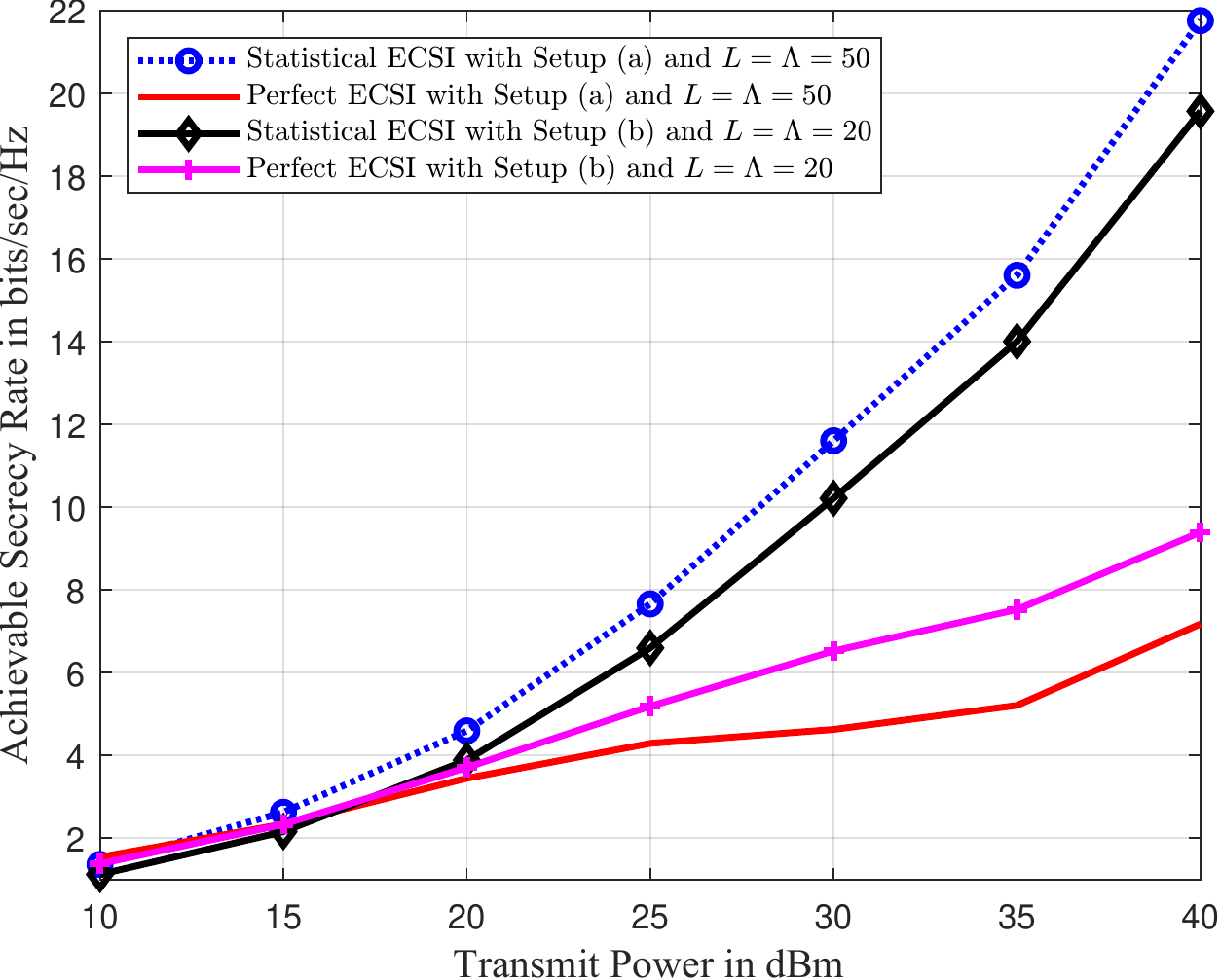}
	\caption{\small{Both PLS system setups with $L = \Lambda$.}}
	\label{fig:Secr_Rates_Equal}
\end{subfigure}
~ 
\,
\begin{subfigure}[h]{0.48\textwidth}
	\includegraphics[width=\textwidth]{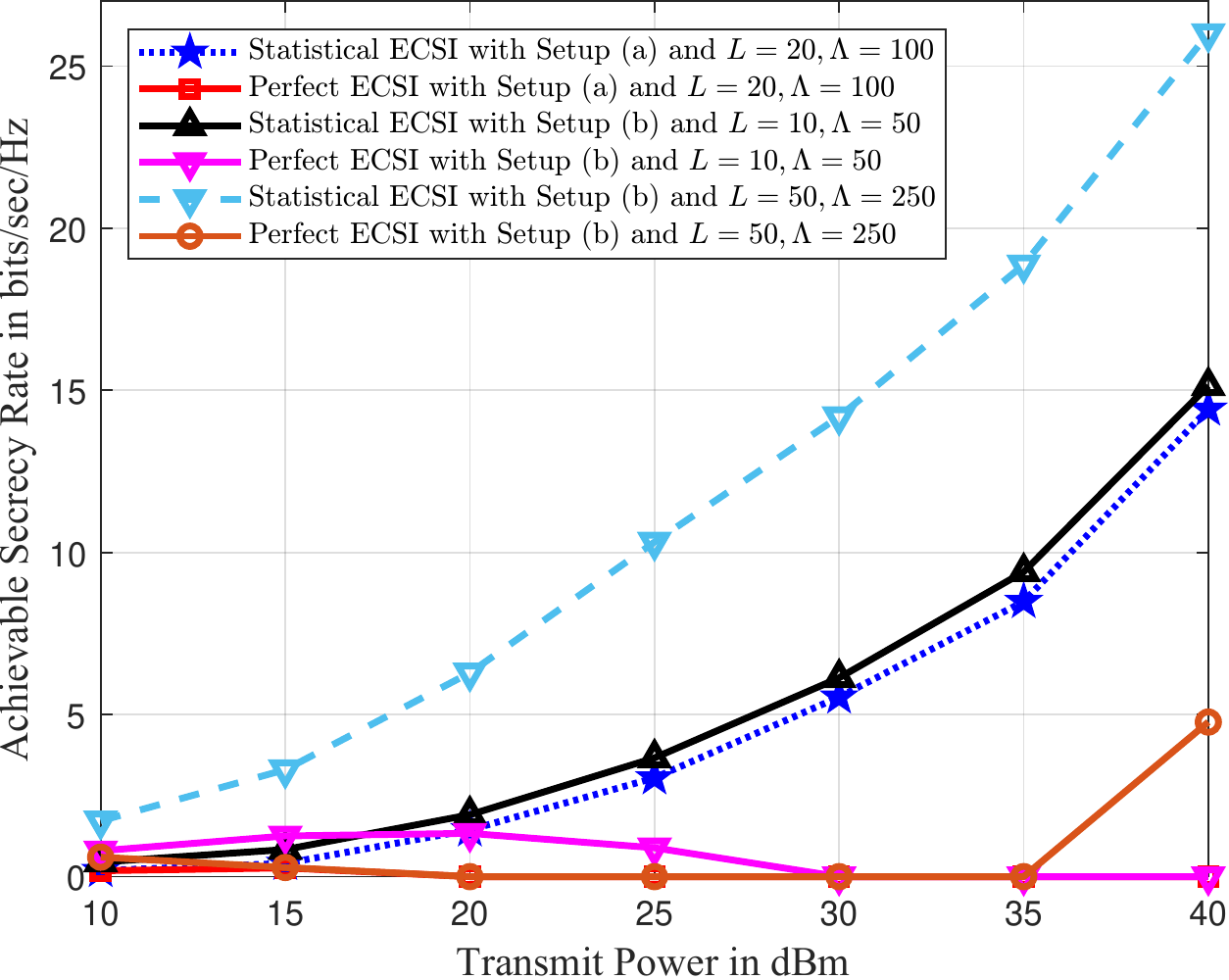}
	\caption{\small{Both PLS system setups with $\Lambda = 5L$.}}
	\label{fig:Secr_Rates_5x}
\end{subfigure}
\caption{\small{Achievable secrecy rates in bits/sec/Hz versus the transmit power $P$ in dBm for both simulated PLS system setups, considering different numbers for $L$ and $\Lambda$. In contrast to Fig$.$~\ref{fig:AN_ONLY_Both}, the legitimate system safeguards communication with BS precoding and AN, RX combining, and RIS reflective beamforming.}}
\label{fig:RIS_RX_Secr_vs_TX_P}
\end{figure} 

In Fig$.$~\ref{fig:RIS_RX_Secr_vs_TX_P}, the achievable secrecy rates for both ECSI knowledge schemes are illustrated as functions of $P$, considering different values for $L$ and $\Lambda$. As depicted in Fig.~\ref{fig:Secr_Rates_Equal} for $L = \Lambda$, all rates follow a non-decreasing trend for increasing values of $P$. Similar to the trend observed in Fig.~\ref{fig:Rates_Full_A}, the statistical knowledge of the eavesdropping channels $\vect{H}_{\rm E}$ and $\vect{G}_{\rm E}$ at the BS, combined with the partial knowledge of $\vect{G}_1$ available at Eve, leads to larger secrecy rates compared to the case of Perfect ECSI, where both the BS and Eve possess all channels perfectly (except of those resulting from the RIS being transparent to them). This witnesses the effectiveness of the proposed scheme, which outperforms that of \cite{Conf} relying on ideal CSI knowledge. The ineffectiveness of the Perfect ECSI scheme is more evident in Fig.~\ref{fig:Secr_Rates_5x}, where the case of a malicious RIS with $500\%$ more reflecting elements than the legitimate one is considered. In this case, the secrecy rates are almost zero for all $P$ values. In the contrary, it is also shown in this figure that, for the considered threat model and the proposed PLS scheme with statistical ECSI, the achievable secrecy rates are large enough to guarantee wireless communication confidentiality for the multiple data streams, even for a small ratio of $L/\Lambda$. 
\begin{figure}[!t]
    \centering
    \begin{subfigure}[h]{0.48\textwidth}
        \includegraphics[width=\textwidth]{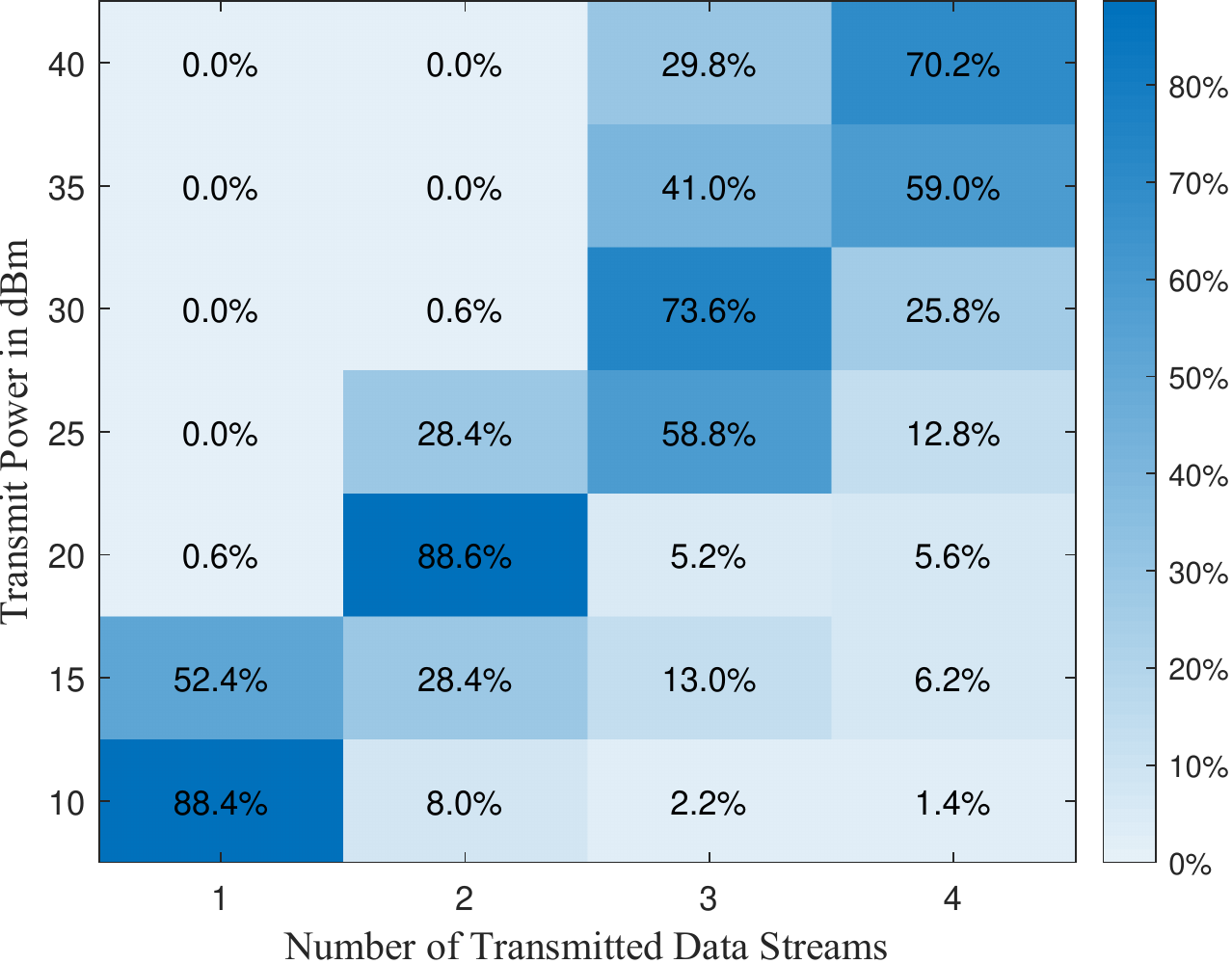}
        \caption{\small{PLS system Setup (a) with $L = 20$ and $\Lambda = 100$.}}
        \label{fig:Heatmap_A}
    \end{subfigure}
    ~ 
			\,
    \begin{subfigure}[h]{0.48\textwidth}
        \includegraphics[width=\textwidth]{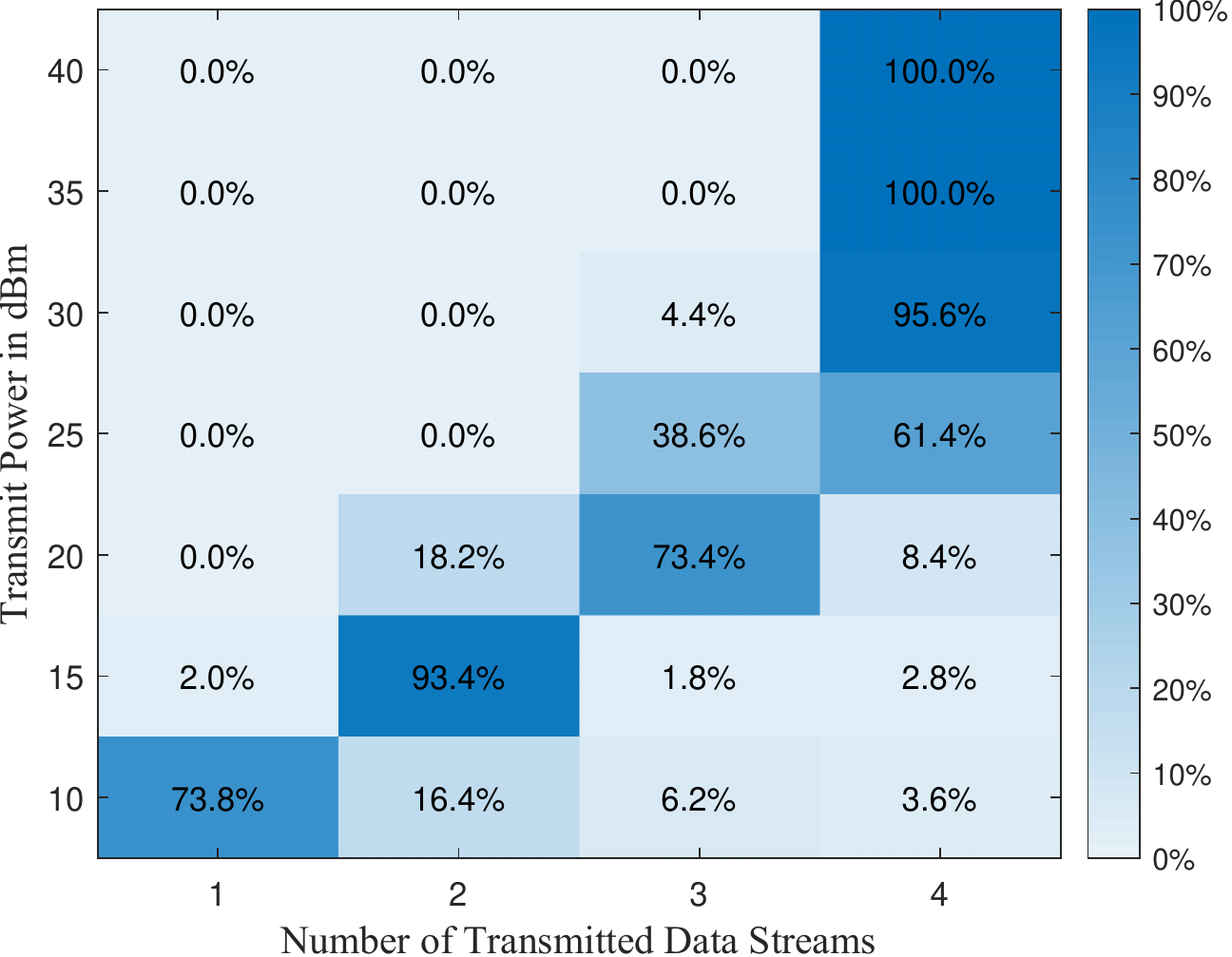}
        \caption{\small{PLS system Setup (b) with $L=\Lambda=20$.}}
        \label{fig:Heatmap_B}
    \end{subfigure}
    \caption{\small{The percentage of the average number of independent data streams versus the transmit power $P$ in dBm for the proposed RIS-empowered MIMO PLS design, considering both simulated setups for different numbers $L$ and $\Lambda$ for the reflecting elements of the legitimate and malicious RISs, respectively. 
		}}
\label{fig:Heatmap_Both}
\end{figure}

The value of $N_d$ referring to the number of the transmitted data streams, as designed by the proposed PLS scheme, for both considered setups is plotted versus $P$ in Fig$.$~\ref{fig:Heatmap_Both} for different numbers $L$ and $\Lambda$ for the elements of the legitimate and malicious RISs, respectively. It can be observed that, for low $P$ values, the designed $N_d$ gets more frequently its lowest possible values (i.e., $N_d=1$ and $2$). For instance, when considering Setup~(a) and $P = 10$ dBm, $N_d=1$ data stream is selected $88.4\%$ times more often than its other three possible values. However, when $P$ increases, larger values for $N_d$ are obtained more frequently. It can be actually seen that the dominating selection frequencies for $N_d$ for all $P$ values in both Figs.~\ref{fig:Heatmap_A} and~\ref{fig:Heatmap_B} exhibit a diagonal pattern. This trend is in agreement with the secrecy waterfilling algorithm \cite{Xing_2016}, which was designed for multi-stream power allocation. Finally, it is demonstrated in the figures that the distribution of the data streams differs between the two considered setups, which apart from the different $\Lambda$ values differ in the number $N$ of BS antenna elements.

\begin{figure}[!t]
\centering
\includegraphics[scale=0.68]{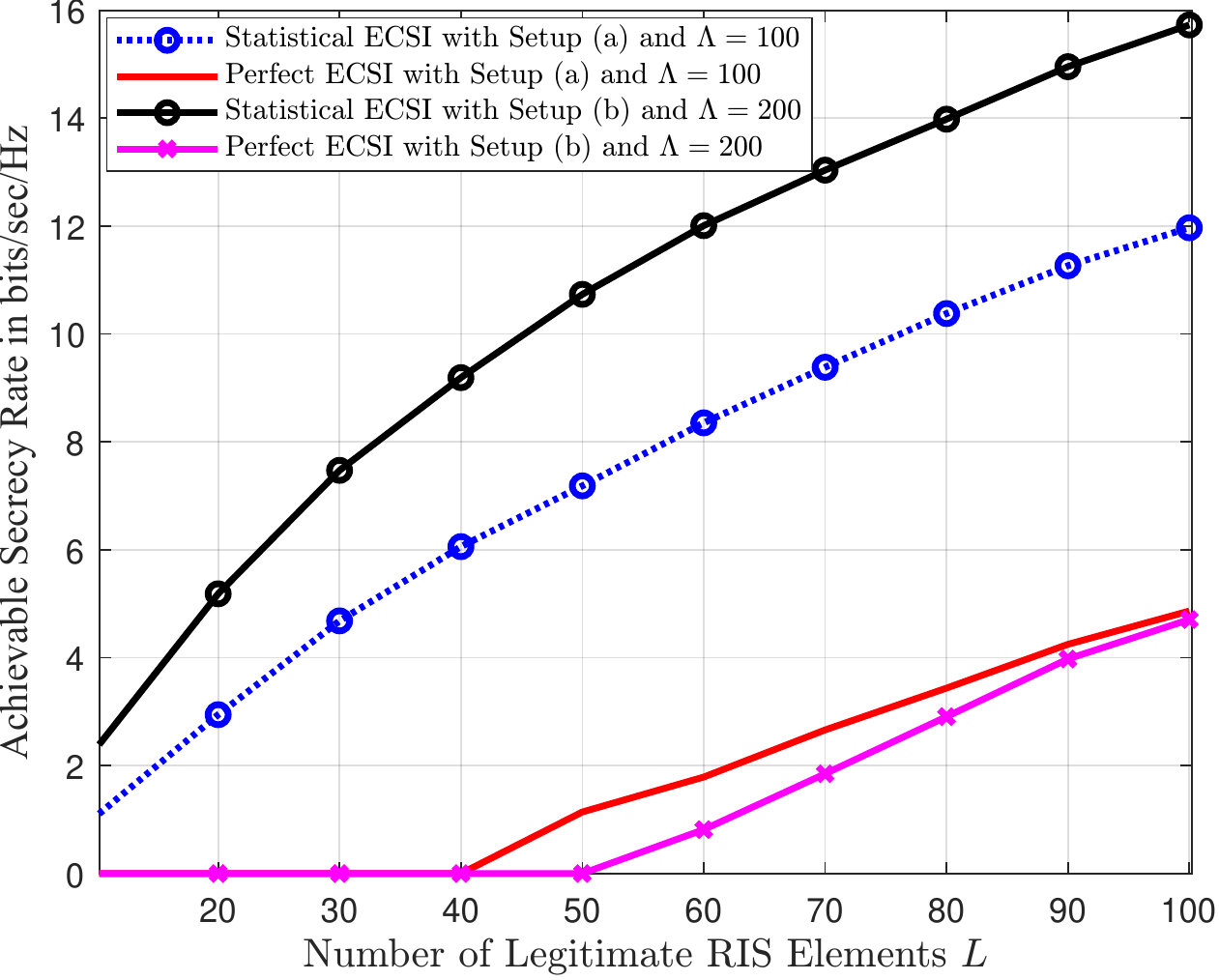}
\caption{\small{Achievable secrecy rates in bits/sec/Hz versus the number $L$ of reflecting elements at the legitimate RIS for both simulated PLS system setups and ECSI models, considering the transmit power $P=25$ dBm and different numbers $\Lambda$ for the reflecting elements at the malicious RIS.}}
\label{fig:RIS_RX_Secr_vs_L}
\end{figure}

The impact of the varying number $L$ of the legitimate RIS's unit elements in the secrecy rate's behavior at $P=25$ dBm is illustrated in Fig$.$~\ref{fig:RIS_RX_Secr_vs_L}. Evidently, for both considered setups and $\Lambda$ values, the secrecy rate exhibits a non-decreasing trend with $L$. This behavior witnesses the secrecy benefits from empowering a legitimate link with an RIS, when this operates in the vicinity of an RIS-empowered eavesdropping system for which the consideration of a malicious RIS is unknown to the legitimate BS. It is noted that almost all investigated $L$ values in the figure are smaller than the considered $\Lambda=\{100,200\}$. In fact, even with a legitimate RIS equipped with $L<\Lambda/5$ elements, non-zero secrecy rates are achievable with the proposed PLS scheme relying on the considered statistical ECSI, while for the Perfect ECSI scheme, it should hold $L\geq\Lambda/2$ for Setup~(a) and $L\geq3\Lambda/10$ for Setup~(b).

\subsection{Area of Influence of the Proposed PLS Scheme} \label{Sec:Numerical_AoI} 
\begin{figure}[!t]
    \centering
    \begin{subfigure}[h]{0.48\textwidth}
        \includegraphics[width=\textwidth]{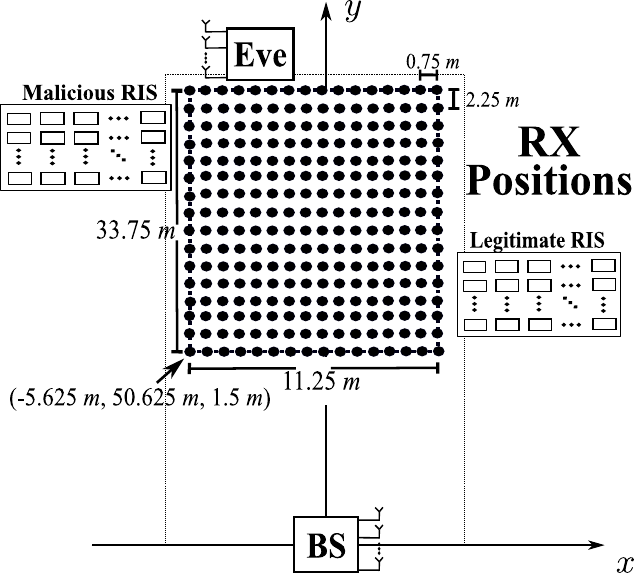}
        \caption{\small{The considered grid of legitimate RX positions.}}
        \label{fig:AoI_noRIS_leg}
    \end{subfigure}
    ~ 
			\,
    \begin{subfigure}[h]{0.48\textwidth}
        \includegraphics[width=\textwidth]{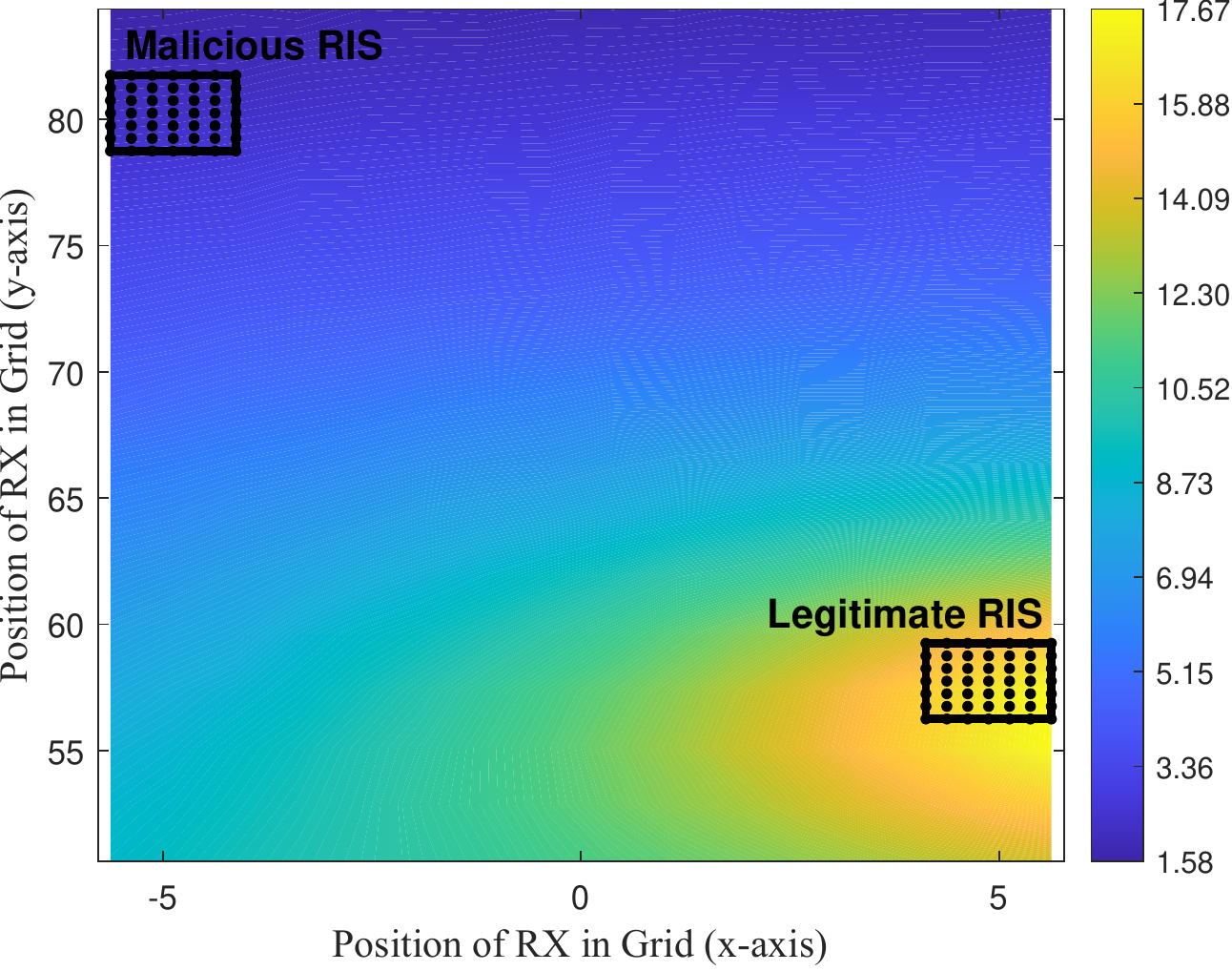}
        \caption{\small{Averaged secrecy rates in bits/sec/Hz for the considered RX positions' grid for $L= 20$ and $\Lambda=100$.}}
        \label{fig:AoI_withRIS_leg}
    \end{subfigure}
    \caption{\small{Area of influence of the proposed RIS-empowered multi-stream MIMO PLS scheme, when operating in the vicinity of an RIS-boosted eavesdropping system, considering a square placement grid of size $11.25 \times 33.75 \,m^2$ for the legitimate RX positions.}}
\label{fig:AoI_both}
\end{figure}

For the specific placements of the nodes in the two simulated setups, the previous figures demonstrated that the proposed RIS-empowered multi-stream MIMO PLS scheme is capable to safeguard legitimate communications even in the presence of an eavesdropper assisted by a large malicious RIS. While this behavior will be similar for various other setups, it is interesting to investigate the secrecy performance distribution of the proposed scheme over a given geographical area. This investigation will unveil the AoI of our scheme, which refers to the identification of the area under which a given secrecy rate threshold can be guaranteed. In Fig$.$~\ref{fig:AoI_both}, we consider a simulation setup similar to Setup~(a), where $\rm RIS_{\rm L}$ is located at $(\frac{w}{2}\,m,\frac{5\ell}{8}\,m,5\,m)$. As shown in Fig.~\ref{fig:AoI_noRIS_leg}, we have considered various possible positions for the legitimate RX within the specified rectangular area, lying in the plane parallel to the $xy$ one and intersecting the $z$-axis at the point $1.5$ $m$. The width and length of this grid area were set equal to $11.25$ and $33.75$ $m$, respectively. The distance between any two consecutive RX positions was selected to be $0.75$ $m$ for the $x$-axis and $2.25$ $m$ for the $y$-axis, which leads to a total number of $16 \times 16 = 256$ different points on the RX placement grid. Moreover, the transmit power was fixed to $P=25$ dBm. For evaluating the AoI of our PLS scheme, we have used the solutions of $\mathcal{OP}_{\rm L}$ and $\mathcal{OP}_{\rm E}$ for each different RX position, considering both the absence and presence of the legitimate RIS.

We have observed that, when the legitimate RIS is absent, the achievable secrecy rates are equal to zero in the sub-grid close to the $100$-element malicious RIS, covering the whole area starting from the point $70.88$ $m$ on the $y$-axis up to the top side of the grid. In addition, in the remaining area, the secrecy rates increase gradually when the RX is located closer to the BS, reaching the maximum value of $4.72$ bits/sec/Hz at the bottom of the RX positions' grid. On the other hand, when a $20$-element legitimate RIS is deployed close to the bottom right corner of the grid and optimized via the proposed scheme, the whole grid area gets boosted in terms of secrecy rate, as demonstrated in Fig$.$~\ref{fig:AoI_withRIS_leg}. In the area where the RX is located close to the malicious RIS (i.e., at the top left corner) and around Eve's position, the secrecy rate values remain still low, but not as low as without the legitimate RIS. For those RX positions, Eve has an eavesdropping advantage due to the smaller distance between itself and the malicious RIS. However, when the RX is placed closer to the legitimate RIS, then a considerable secrecy rate improvement is observed. Specifically, in the sub-grid around the bottom right corner, which covers an area of about $7.5 \times 13.5 = 101.25$ $m^2$, the achievable secrecy rate values are significantly improved, creating an area of secrecy rate boosting. In particular, the rates in this sub-grid reach at least the $62 \%$ of the maximum value $17.67$ bps/Hz and at least the $200\%$ of their achievable values for the case where the legitimate RIS is absent. Evidently, an RIS creates an AoI close to it, which actually holds for both the legitimate and eavesdropping sides.

\section{Conclusions} \label{Sec:Conclusion} 
In this paper, we proposed and studied an RIS-empowered multi-stream MIMO PLS communication system, where RISs were deployed from both the legitimate and the eavesdropping sides, while their presence was transparent between the competitive systems. We focused on the case where the malicious RIS is placed close to the eavesdropper, whereas the legitimate RIS is located either close to the legitimate BS or RX. A novel threat model for RIS-boosted eavesdropping systems was proposed, for which we designed the eavesdropper’s receive combining matrix and the reflection coefficients of the malicious RIS. We presented an ergodic secrecy rate maximization algorithm for the joint design of the BS precoding matrix and number of data streams, the AN covariance matrix, the RX combining matrix, and the RIS reflection coefficients for the legitimate system, proving also its convergence to a stationary point and analyzing its computational requirements. Our extensive performance evaluation results showcased that the proposed RIS-empowered PLS scheme is capable of safeguarding multi-stream MIMO communication over RIS-boosted eavesdropping systems with much larger RISs. We also quantified the geographical area of secrecy rate boosting offered by the proposed scheme for an example system setup, which was shown to be close to the legitimate RIS, while depending on its relative size with respect to the malicious RIS.

\appendices
\section{Proof of Proposition \ref{Prop:Prop_Ergodic_R_E}} \label{appx:Prop_Ergodic_R_E} 
To derive an upper bound for $\mathbb{E}_{\vect{H}_{\rm E},\vect{G}_{\rm E}}[\hat{\mathcal{R}}_{\rm E,inst}]$ (we next omit the indices in the expectations, implying averaging over all involved random matrices), we use of the following lemma \cite{Wu_2014}.
\begin{Lem} \label{Lemma_Exp_Bound}
	Let $\vect{Y} \succeq \vect{0}$ be an $n \times n$ random matrix with $\mathbb{E}[\vect{Y}] = \vect{M}_y$. Let also $\vect{A}, \vect{B} \succ \vect{0}$ be $n \times n$ constant matrices with $\vect{A} - \vect{B} \succeq \vect{0}$. Then, the following inequality holds:
	\begin{equation*}
		\mathbb{E}[\log_2\left\lvert \vect{I}_N + \vect{Y}\vect{A} \right\rvert] - \mathbb{E}[\log_2\left\lvert \vect{I}_N + \vect{Y}\vect{B} \right\rvert] \leq \log_2\left\lvert \vect{I}_N + \vect{M}_y\vect{A} \right\rvert - \log_2\left\lvert \vect{I}_N + \vect{M}_y\vect{B} \right\rvert.
	\end{equation*}
\end{Lem}

Using the Sylvester's determinant identity, $\hat{\mathcal{R}}_{\rm E,inst}$ can be decomposed as $\hat{\mathcal{R}}_{\rm E,inst} = - \hat{\mathcal{R}}_{\rm E_1,inst} + \hat{\mathcal{R}}_{\rm E_2,inst}$, where $\hat{\mathcal{R}}_{\rm E_1,inst} \triangleq \log_2\left\lvert \vect{I}_N + \sigma^{-2} \hat{\vect{H}}_{\rm E}^H \hat{\vect{H}}_{\rm E} \vect{Z} \right\rvert$ and $\hat{\mathcal{R}}_{\rm E_2,inst} \ \log_2\left\lvert \vect{I}_N + \sigma^{-2} \hat{\vect{H}}_{\rm E}^H \hat{\vect{H}}_{\rm E} \bar{\vect{X}} \right\rvert$. It clearly holds that $\bar{\vect{X}} - \vect{Z} \succeq \vect{0}$, hence, the assumptions of the latter lemma are satisfied. This deduces to $\mathbb{E}[\hat{\mathcal{R}}_{\rm E,inst}] \leq \log_2\left\lvert \vect{I}_N + \sigma^{-2} \mathbb{E}[\hat{\vect{H}}_{\rm E}^H \hat{\vect{H}}_{\rm E}] \bar{\vect{X}} \right\rvert - \log_2\left\lvert \vect{I}_N + \sigma^{-2} \mathbb{E}[\hat{\vect{H}}_{\rm E}^H \hat{\vect{H}}_{\rm E}] \vect{Z} \right\rvert$. Then, unfolding the term $\mathbb{E}[\hat{\vect{H}}_{\rm E}^H \hat{\vect{H}}_{\rm E}]$ according to the definition of $\hat{\vect{H}}_{\rm E}$, yields the following expression:
\begin{equation*}
	\mathbb{E}[\hat{\vect{H}}_{\rm E}^H \hat{\vect{H}}_{\rm E}] = \mathbb{E}[\vect{H}_{\rm E}^H \vect{H}_{\rm E}] + \mathbb{E}[\vect{H}_{\rm E}^H \vect{G}_{\rm E} \vect{\Phi} \vect{H}_1] + \mathbb{E}[\vect{H}_1^H \vect{\Phi}^H \vect{G}_{\rm E}^H \vect{H}_{\rm E}] + \mathbb{E}[\vect{H}_1^H \vect{\Phi}^H \vect{G}_{\rm E}^H \vect{G}_{\rm E} \vect{\Phi} \vect{H}_1].
\end{equation*}
The second and third terms in the right-hand side of this expression are equal to zero, since the direct channel $\vect{H}_{\rm E}$ and $\vect{G}_{\rm E}$, included in the RIS-parameterized cascaded channel, are independent. By using the matrix definitions in the Proposition's statement, the proof is complete.

\section{Proof of Corollary \ref{Thm:OP_L_Find_kappa}} \label{appx:Thm_OP_L_Find_kappa} 
We first replace \eqref{eq:optimal_U} and the expression for $\operatorname{vec}(\vect{U}_\kappa^{\star})$ into the following trace operator that appears in the constraint for the RX combining matrix:
\begin{align}
\trace(\vect{U}_{\rm opt}^H(\kappa) \vect{U}_{\rm opt}(\kappa)) &= \trace\left( (\vect{I}_{N_d} \otimes \operatorname{vec}(\vect{U}_\kappa^{\star})^H)(\mathbf{Q}_{N_d} \otimes \vect{I}_M)(\vect{I}_{N_d}\otimes \operatorname{vec}(\vect{U}_\kappa^{\star})) \right) \label{eq:kappa_s1} \\
&= \trace\left( (\mathbf{Q}_{N_d} \otimes \vect{I}_M)(\vect{I}_{N_d}\otimes \operatorname{vec}(\vect{U}_\kappa^{\star})) (\vect{I}_{N_d} \otimes \operatorname{vec}(\vect{U}_\kappa^{\star})^H) \right) \label{eq:kappa_s2} \\
&= \trace\left( (\mathbf{Q}_{N_d} \otimes \vect{I}_M)(\vect{I}_{N_d} \otimes \operatorname{vec}(\vect{U}_\kappa^{\star})\operatorname{vec}(\vect{U}_\kappa^{\star})^H ) \right) \label{eq:kappa_s3} \\
&= \trace\left( \mathbf{Q}_{N_d} \otimes \operatorname{vec}(\vect{U}_\kappa^{\star})\operatorname{vec}(\vect{U}_\kappa^{\star})^H \right) \label{eq:kappa_s4} \\
&= \trace\left( \mathbf{Q}_{N_d} \right) \trace\left( \operatorname{vec}(\vect{U}_\kappa^{\star})\operatorname{vec}(\vect{U}_\kappa^{\star})^H \right) \label{eq:kappa_s5} \\
&= N_d \trace\left( (\vect{F}^T \otimes \vect{E} + \kappa\vect{I}_{M N_d} )^{-2} \operatorname{vec}(\vect{J}) \operatorname{vec}(\vect{J})^H \right), \label{eq:kappa_s6}
\end{align}
where $\mathbf{Q}_{N_d}\triangleq\operatorname{vec}(\vect{I}_{N_d})\operatorname{vec}(\vect{I}_{N_d})^T$. In this derivation, we first applied the mixed-product property $(\vect{A} \otimes \vect{B})(\vect{C} \otimes \vect{D}) = (\vect{A} \vect{C}) \otimes (\vect{B} \vect{D})$ in \eqref{eq:kappa_s1}, followed by trace's cyclic property in \eqref{eq:kappa_s2}. The mixed-product property was re-applied in \eqref{eq:kappa_s3}, the identity $\vect{A} \otimes \vect{B} = (\vect{A} \otimes \vect{I}_1)(\vect{I}_2 \otimes \vect{B})$ was invoked in \eqref{eq:kappa_s4}, and the identity $\trace(\vect{A} \otimes \vect{B}) = \trace(\vect{A}) \trace(\vect{B})$ was used in \eqref{eq:kappa_s5}. Finally, \eqref{eq:kappa_s6} is deduced from the fact that $\trace\left(\mathbf{Q}_{N_d}\right)=N_d$ 
and the replacement of $\operatorname{vec}(\vect{U}_\kappa^{\star})$. 
By applying the eigenvalue decomposition $\vect{F}^T \otimes \vect{E} = \bar{\vect{Q}} \vect{\Xi} \bar{\vect{Q}}^H$ (as in the Corollary's statement) in the Kronecker product inside the expression for $\operatorname{vec}(\vect{U}_\kappa^{\star})$, it can be readily verified after some algebraic manipulations that $\kappa^{\star}$ is derived as described in \eqref{eq:optimum_kappa}; the proof is, thus, complete.

\section{Proof of Theorem \ref{Thm:OP_L_KKT_Convergence}} \label{appx:Thm_OP_L_KKT_Convergence} 
For each $p$-th iteration in the inner loop of Algorithm~\ref{alg:OP_L_Overall_Algorithm} and for each feasible value of $m$ for the number of independent data streams, it holds that: 
\begin{align}
&\hat{\mathcal{R}}_{\rm s}^{\rm lb}\left(\vect{U}^{(p)}_m, \vect{V}^{(p)}_m,\vect{Z}^{(p)}_m,\vect{\phi}^{(p)}_m\right) = \bar{\mathcal{R}}_{\rm s}^{\rm lb}\left(\{\vect{A}_i^{(p)}\}_{i=1}^2,\{\vect{S}_j^{(p)}\}_{j=1}^3,\vect{U}^{(p)}_m,\vect{V}^{(p)}_m,\vect{Z}^{(p)}_m,\vect{\phi}^{(p)}_m\right)\nonumber \\
 &\leq \bar{\mathcal{R}}_{\rm s}^{\rm lb}\left(\{\vect{A}_i^{(p+1)}\}_{i=1}^2,\{\vect{S}_j^{(p+1)}\}_{j=1}^3,\vect{U}^{(p+1)}_m,\vect{V}^{(p+1)}_m,\vect{Z}^{(p+1)}_m,\vect{\phi}^{(p)}_m\right) = g(\vect{\phi}^{(p)}_m) \leq g(\vect{\phi}^{(p+1)}_m)\nonumber \\
 &\leq \bar{\mathcal{R}}_{\rm s}^{\rm lb}\left(\{\vect{A}_i^{(p+1)}\}_{i=1}^2,\{\vect{S}_j^{(p+1)}\}_{j=1}^3,\vect{U}^{(p+1)}_m,\vect{V}^{(p+1)}_m,\vect{Z}^{(p+1)}_m,\vect{\phi}^{(p+1)}_m\right)\nonumber \\
 &= \hat{\mathcal{R}}_{\rm s}^{\rm lb}\left(\vect{U}^{(p+1)}_m, \vect{V}^{(p+1)}_m,\vect{Z}^{(p+1)}_m,\vect{\phi}^{(p+1)}_m\right),\nonumber
\end{align}
where the first and the last equality result from Lemma~\ref{Lemma_AO_AN}, the first inequality follows from the algorithmic steps $4$--$8$, and the equality (as well as the inequalities) related to the function $g(\cdot)$ hold by the MM approximation principle. Note that the auxiliary matrix variables refer to the specific $m$-th value of the outer loop. This proves the non-decreasing trend of the algorithm.

To prove the satisfaction of the KKT conditions at $I$ iterations of the inner loop in Algorithm~\ref{alg:OP_L_Overall_Algorithm}, we begin by observing that $\vect{U}^{(I)}_m$ satisfies the KKT conditions of the maximization problem with the objective function $\bar{\mathcal{C}}_{\rm s}(\vect{U}_m,\vect{V}^{(I)}_m,\vect{Z}^{(I)}_m,\vect{\phi}^{(I)}_m) \triangleq -\trace(\vect{S}_1^{(I)} \vect{M}_1^{(I)}(\vect{U}_m))$ \cite{Shi_2015}, where $\vect{S}_1^{(I)}$ and $\vect{M}_1^{(I)}(\vect{U}_m)$ refer to the specific $m$-th value of the outer loop, with the latter depending on $\vect{U}_m$ and $\vect{A}_1^{(I)}$. The Lagrangian of this function is given similar to \eqref{eq:Lagrangian_U} by the following expression (we henceforth omit the subscript $m$ for clarity):
\begin{equation*}
 \mathcal{L}_{\bar{\mathcal{C}}_{\rm s}}(\vect{U},\vect{V}^{(I)},\vect{Z}^{(I)},\vect{\phi}^{(I)}) = \bar{\mathcal{C}}_{\rm s}(\vect{U},\vect{V}^{(I)},\vect{Z}^{(I)},\vect{\phi}^{(I)}) - \kappa (\trace(\vect{U}^H \vect{U}) - 1),
\end{equation*}
which evaluated at the point $\vect{U}^{(I)}$ and for the optimal Lagrange multiplier yields
\begin{equation*}
 \nabla_{\vect{U}} \mathcal{L}_{\bar{\mathcal{C}}_{\rm s}}(\vect{U},\vect{V}^{(I)},\vect{Z}^{(I)},\vect{\phi}^{(I)}) \Bigr\rvert_{\vect{U}\! =\! \vect{U}^{(I)}} = \nabla_{\vect{U}} \bar{\mathcal{C}}_{\rm s}(\vect{U},\vect{V}^{(I)},\vect{Z}^{(I)},\vect{\phi}^{(I)}) \Bigr\rvert_{\vect{U} = \vect{U}^{(I)}} - \kappa^{\star} \vect{U}^{(I)} = \vect{0}_{M\times N_d}.
\end{equation*}
Then, it can be easily shown for $\bar{\mathcal{R}}_{\rm s}^{\rm lb}\left(\cdot\right)$'s gradient that:
\begin{equation*}
\begin{aligned}
 \nabla_{\vect{U}} \bar{\mathcal{R}}_{\rm s}^{\rm lb}\left(\{\vect{A}_i^{(I)}\}_{i=1}^2,\{\vect{S}_j^{(I)}\}_{j=1}^3,\vect{U},\vect{V}^{(I)},\vect{Z}^{(I)},\vect{\phi}^{(I)}\right) \Bigr\rvert_{\vect{U} = \vect{U}^{(I)}} \!=\! \nabla_{\vect{U}} \bar{\mathcal{C}}_{\rm s}(\vect{U},\vect{V}^{(I)},\vect{Z}^{(I)},\vect{\phi}^{(I)}) \Bigr\rvert_{\vect{U} = \vect{U}^{(I)}},
 \end{aligned}
\end{equation*}
which can be used for the following derivations:
\begin{align}
&\nabla_{\vect{U}} \bar{\mathcal{R}}_{\rm s}^{\rm lb}\left(\{\vect{A}_i^{(I)}\}_{i=1}^2,\{\vect{S}_j^{(I)}\}_{j=1}^3,\vect{U},\vect{V}^{(I)},\vect{Z}^{(I)},\vect{\phi}^{(I)}\right) \Bigr\rvert_{\vect{U} = \vect{U}^{(I)}}
 \stackrel{(a)}{=} - \trace\left( \vect{S}_1^{(I)} \nabla_{\vect{U}} \vect{M}_1^{(I)}(\vect{U}) \right) \Bigr\rvert_{\vect{U} = \vect{U}^{(I)}}\nonumber \\
 &\stackrel{(b)}{=} -\trace \left( (\vect{M}_1^{(I)}(\vect{U}))^{-1} \nabla_{\vect{U}} \vect{M}_1^{(I)}(\vect{U}) \right) \Bigr\rvert_{\vect{U} = \vect{U}^{(I)}} 
 \stackrel{(c)}{=} - \nabla_{\vect{U}} \log \left\lvert \vect{M}_1^{(I)}(\vect{U}) \right\rvert_{\vect{U} = \vect{U}^{(I)}}\nonumber \\
&\stackrel{(d)}{=} \nabla_{\vect{U}} \log \left\lvert \vect{S}_1^{(I)}(\vect{U}) \right\rvert_{\vect{U} = \vect{U}^{(I)}} \stackrel{(e)}{=} \nabla_{\vect{U}} \hat{\mathcal{R}}_{\rm s}^{\rm lb}(\vect{U}, \vect{V}^{(I)},\vect{Z}^{(I)},\vect{\phi}^{(I)})\Bigr\rvert_{\vect{U} = \vect{U}^{(I)}},\nonumber
\end{align}
where $(a)$ follows from the linearity of the gradient and trace operators and by replacing the optimal expressions of $\vect{S}_1^{(I)}$ and $\vect{A}_1^{(I)}$, $(b)$ and $(d)$ are due to Lemma~\ref{Lemma_AO_AN}, $(c)$ results from the matrix differentiation identity $\rm d \log \det (\vect{X}) = \trace(\vect{X}^{-1} \rm d \vect{X})$, and $(e)$ yields from \eqref{eq:optimal_S_1}. In a similar manner, it can be verified that $\vect{V}^{(I)}$ and $\vect{Z}^{(I)}$ constitute a KKT point. Also, it holds from \cite[Th. 4.3.1]{Absil_2008} that Algorithm~\ref{alg:OP_L_phi} converges to a point where the gradient of $g(\vect{\phi}|\tilde{\vect{\phi}})$ is zero. Finally, it holds by the MM principle that $\nabla_{\vect{\phi}} \hat{\mathcal{R}}_{\rm s}^{\rm lb}(\vect{\phi})\Bigr\rvert_{\vect{\phi} = \vect{\phi}^{(I)}} = \nabla_{\vect{\phi}} g(\vect{\phi}|\tilde{\vect{\phi}})\Bigr\rvert_{\vect{\phi} = \vect{\phi}^{(I)}}$. Following analogous steps to those for $\vect{U}^{(I)}$, it results that $\vect{\phi}^{(I)}$ satisfies the KKT conditions. This concludes the proof. 

\bibliographystyle{IEEEtran}
\bibliography{references_new}

\begin{thebibliography}{10}
\providecommand{\url}[1]{#1}
\csname url@samestyle\endcsname
\providecommand{\newblock}{\relax}
\providecommand{\bibinfo}[2]{#2}
\providecommand{\BIBentrySTDinterwordspacing}{\spaceskip=0pt\relax}
\providecommand{\BIBentryALTinterwordstretchfactor}{4}
\providecommand{\BIBentryALTinterwordspacing}{\spaceskip=\fontdimen2\font plus
\BIBentryALTinterwordstretchfactor\fontdimen3\font minus
  \fontdimen4\font\relax}
\providecommand{\BIBforeignlanguage}[2]{{%
\expandafter\ifx\csname l@#1\endcsname\relax
\typeout{** WARNING: IEEEtran.bst: No hyphenation pattern has been}%
\typeout{** loaded for the language `#1'. Using the pattern for}%
\typeout{** the default language instead.}%
\else
\language=\csname l@#1\endcsname
\fi
#2}}
\providecommand{\BIBdecl}{\relax}
\BIBdecl

\bibitem{Conf}
G.~C. Alexandropoulos \emph{et~al.}, ``Safeguarding {MIMO} communications with
  reconfigurable metasurfaces and artificial noise,'' in \emph{Proc. IEEE ICC},
  Montreal Canada, Jun. 2021, pp. 1--6.

\bibitem{George_RIS_TWC2019}
C.~Huang \emph{et~al.}, ``{Reconfigurable intelligent surfaces for energy
  efficiency in wireless communication},'' \emph{IEEE Trans. Wireless Commun.},
  vol.~18, no.~8, pp. 4157--4170, Aug. 2019.

\bibitem{Wu_RIS_TWC2019}
Q.~Wu and R.~Zhang, ``{Intelligent reflecting surface enhanced wireless network
  via joint active and passive beamforming},'' \emph{IEEE Trans. Wireless
  Commun.}, vol.~18, no.~11, pp. 5394--5409, Nov. 2019.

\bibitem{rise6g}
E.~Calvanese~Strinati \emph{et~al.}, ``Wireless environment as a service
  enabled by reconfigurable intelligent surfaces: {T}he {RISE-6G}
  perspective,'' in \emph{Proc. Joint EuCNC \& 6G Summit}, Porto, Portugal,
  Jun. 2021, pp. 1--6.

\bibitem{Kaina_metasurfaces_2014}
N.~Kaina \emph{et~al.}, ``Shaping complex microwave fields in reverberating
  media with binary tunable metasurfaces,'' \emph{Sci. Rep. 4}, pp. 1--7,
  Article No 076401, 2014.

\bibitem{Marco_Visionary_2019}
M.~D. Renzo \emph{et~al.}, ``{Smart radio environments empowered by AI
  reconfigurable meta-surfaces: An idea whose time has come},'' \emph{EURASIP
  J. Wireless Commun. Netw.}, vol. 129, pp. 1--20, May 2019.

\bibitem{HMIMO}
C.~Huang \emph{et~al.}, ``Holographic {MIMO} surfaces for {6G} wireless
  networks: {O}pportunities, challenges, and trends,'' \emph{IEEE Wireless
  Commun.}, vol.~27, no.~5, pp. 118--125, Oct. 2020.

\bibitem{Tsinghua_RIS_Tutorial}
M.~Jian \emph{et~al.}, ``Reconfigurable intelligent surfaces for wireless
  communications: {O}verview of hardware designs, channel models, and
  estimation techniques,'' \emph{Intell. Converged Netw.}, vol.~3, no.~1, pp.
  1--32, Mar. 2022.

\bibitem{Yang_ComMag_2015}
N.~Yang \emph{et~al.}, ``Safeguarding {5G} wireless communication networks
  using physical layer security,'' \emph{IEEE Commun. Mag.}, vol.~53, no.~4,
  pp. 20--27, Apr. 2015.

\bibitem{Mensi_2021}
N.~Mensi \emph{et~al.}, ``Physical layer security for v2i communications:
  Reflecting surfaces vs. relaying,'' in \emph{Proc. IEEE GLOBECOM}, Dec. 2021,
  pp. 1--6.

\bibitem{Chen_2019}
J.~Chen \emph{et~al.}, ``Intelligent reflecting surface: {A} programmable
  wireless environment for physical layer security,'' \emph{IEEE Access},
  vol.~7, pp. 82\,599--82\,612, Jul. 2019.

\bibitem{Hong_2020}
S.~Hong \emph{et~al.}, ``Artificial-noise-aided secure {MIMO} wireless
  communications via intelligent reflecting surface,'' \emph{IEEE Trans.
  Commun.}, vol.~68, no.~12, pp. 7851--7866, Dec. 2020.

\bibitem{Dong_2020b}
L.~{Dong} and H.~M. {Wang}, ``Enhancing secure {MIMO} transmission via
  intelligent reflecting surface,'' \emph{IEEE Trans. Wireless Commun},
  vol.~19, no.~11, pp. 7543--7556, Nov. 2020.

\bibitem{Feng_2019}
B.~Feng \emph{et~al.}, ``Secure transmission strategy for intelligent
  reflecting surface enhanced wireless system,'' in \emph{Proc. IEEE WCSP},
  Xi'an, China, Oct. 2019, pp. 1--6.

\bibitem{Wang_2020b}
H.-M. Wang \emph{et~al.}, ``Intelligent reflecting surfaces assisted secure
  transmission without eavesdropper's {CSI},'' \emph{IEEE Signal Process.
  Lett.}, vol.~27, pp. 1300--1304, Jul. 2020.

\bibitem{Hong_2021b}
S.~Hong \emph{et~al.}, ``Robust transmission design for intelligent reflecting
  surface-aided secure communication systems with imperfect cascaded {CSI},''
  \emph{IEEE Trans. Wirel. Commun.}, vol.~20, no.~4, pp. 2487--2501, Apr. 2021.

\bibitem{Liu_2021f}
C.~Liu \emph{et~al.}, ``{RIS}-assisted secure transmission exploiting
  statistical {CSI} of eavesdropper,'' in \emph{Proc. IEEE GLOBECOM}, Madrid,
  Spain, Dec. 2021, pp. 1--6.

\bibitem{Li_2022c}
Z.~Li \emph{et~al.}, ``Secure multicast energy-efficiency maximization with
  massive {RIS}s and uncertain {CSI}: First-order algorithms and convergence
  analysis,'' \emph{IEEE Trans. Wirel. Commun.}, early access, Feb. 2022.

\bibitem{Shu_2021b}
F.~Shu \emph{et~al.}, ``Enhanced secrecy rate maximization for directional
  modulation networks via {IRS},'' \emph{IEEE Trans. Commun.}, vol.~69, no.~12,
  pp. 8388--8401, Dec. 2021.

\bibitem{Lyu_2020}
B.~{Lyu} \emph{et~al.}, ``{IRS}-based wireless jamming attacks: When jammers
  can attack without power,'' \emph{IEEE Wireless Commun. Lett.}, vol.~9,
  no.~10, pp. 1663--1667, Oct. 2020.

\bibitem{Zheng_2020}
X.~Zheng \emph{et~al.}, ``Uplink channel estimation and signal extraction
  against malicious {IRS} in massive {MIMO} system,'' in \emph{Proc. IEEE ICC},
  Montreal, Canada, Jun. 2021, pp. 1--6.

\bibitem{Huang_2021}
K.-W. Huang and H.-M. Wang, ``Intelligent reflecting surface aided pilot
  contamination attack and its countermeasure,'' \emph{IEEE Trans. Wireless
  Commun.}, vol.~20, no.~1, pp. 345--359, Jan. 2021.

\bibitem{Wang_2021}
J.~Wang \emph{et~al.}, ``Joint transmit beamforming and phase shift design for
  reconfigurable intelligent surface assisted {MIMO} systems,'' \emph{IEEE
  Trans. Cogn. Commun. Netw.}, vol.~7, no.~2, pp. 354--368, Jun. 2021.

\bibitem{George_RIS_2020}
G.~C. Alexandropoulos and E.~Vlachos, ``A hardware architecture for
  reconfigurable intelligent surfaces with minimal active elements for explicit
  channel estimation,'' in \emph{Proc. IEEE ICASSP}, Barcelona, Spain, May
  2020, pp. 9175--9179.

\bibitem{HRIS_Mag}
G.~C. Alexandropoulos \emph{et~al.}, ``Hybrid reconfigurable intelligent
  metasurfaces: {E}nabling simultaneous tunable reflections and sensing for
  {6G} wireless communications,'' 2021, [Online]
  https://arxiv.org/abs/2104.04690.

\bibitem{Liu_2017}
Y.~Liu \emph{et~al.}, ``Securing relay networks with artificial noise: {A}n
  error performance based approach,'' \emph{MDPI Entropy 19}, no. 8: 384, pp.
  7495--7505, Jul. 2017.

\bibitem{Oggier_2011}
F.~{Oggier} and B.~{Hassibi}, ``The secrecy capacity of the {MIMO} wiretap
  channel,'' \emph{IEEE Trans. Inf. Theory}, vol.~57, no.~8, pp. 4961--4972,
  Aug. 2011.

\bibitem{Niu_2022}
H.~Niu \emph{et~al.}, ``Artificial noise elimination: From the perspective of
  eavesdroppers,'' \emph{IEEE Trans. Commun.}, vol.~70, no.~7, pp. 4745--4754,
  Jul. 2022.

\bibitem{Shi_2015}
Q.~Shi \emph{et~al.}, ``Secure beamforming for {MIMO} broadcasting with
  wireless information and power transfer,'' \emph{IEEE Trans. Wireless
  Commun.}, vol.~14, no.~5, pp. 2841--2853, May 2015.

\bibitem{Boyd_2004}
S.~Boyd and L.~Vandenberghe, \emph{Convex Optimization}.\hskip 1em plus 0.5em
  minus 0.4em\relax Cambridge University Press, 2004.

\bibitem{sun2016majorization}
Y.~Sun \emph{et~al.}, ``Majorization-minimization algorithms in signal
  processing, communications, and machine learning,'' \emph{IEEE Trans. Signal
  Process.}, vol.~65, no.~3, pp. 794--816, Feb. 2016.

\bibitem{Zhang_2017}
X.-D. Zhang, \emph{Matrix Analysis and Applications}.\hskip 1em plus 0.5em
  minus 0.4em\relax Cambridge University Press, 2017.

\bibitem{Absil_2008}
P.-A. Absil \emph{et~al.}, \emph{Optimization Algorithms on Matrix
  Manifolds}.\hskip 1em plus 0.5em minus 0.4em\relax Princeton University
  Press, 2008.

\bibitem{Shewchuk_1994}
J.~R. Shewchuk, ``An introduction to the conjugate gradient method without the
  agonizing pain,'' USA, Tech. Rep., 1994.

\bibitem{STAR_RIS_2022}
J.~He \emph{et~al.}, ``Simultaneous indoor and outdoor {3D} localization with
  {STAR-RIS}-assisted millimeter wave systems,'' in \emph{Proc. IEEE VTC-Fall},
  London/Beijing, UK/China, Sep. 2022, pp. 1--6.

\bibitem{gupta2018matrix}
A.~Gupta and D.~Nagar, \emph{Matrix variate distributions}.\hskip 1em plus
  0.5em minus 0.4em\relax USA: Chapman and Hall, 2018, vol. 104.

\bibitem{Xing_2016}
H.~{Xing} \emph{et~al.}, ``Secrecy wireless information and power transfer in
  fading wiretap channel,'' \emph{IEEE Trans. Veh. Technol.}, vol.~65, no.~1,
  pp. 180--190, Jan. 2016.

\bibitem{Wu_2014}
Y.~Wu \emph{et~al.}, ``Transmit designs for the {MIMO} broadcast channel with
  statistical {CSI},'' \emph{IEEE Trans. Signal Process.}, vol.~62, no.~17, pp.
  4451--4466, Sep. 2014.

\end{thebibliography}
\end{document}